\documentclass[manuscript,screen,authordraft,nonacm,review=false,timestamp=false]{acmart}

\usepackage[yyyymmdd]{datetime}

\usepackage{tabularx}
\usepackage{siunitx}
\usepackage{arydshln}
\usepackage{xcolor}
\usepackage{dcolumn} 
\newcolumntype{d}[1]{D{.}{.}{#1}}
\usepackage{enumitem}

\usepackage{subcaption}

\newcommand{\osfrepository}{\href{https://osf.io/svax9/?view_only=cd176c08c2d4422baabf6bf815018bd7}{https://osf.io/svax9/?view\_only=cd176c08c2d4422baabf6bf815018bd7}}

\AtBeginDocument{%
  }

\setcopyright{acmlicensed}
\copyrightyear{2025}
\acmYear{2025}
\acmDOI{XXXXXXX.XXXXXXX}

\acmConference[CHI '25]{Proceedings of the CHI Conference on Human Factors in Computing Systems}{April 26 -- May 1, 2025}{Yokohama, Japan}
\acmBooktitle{Proceedings of the CHI Conference on Human Factors in Computing Systems (CHI '25), April 26 -- May 1, 2025, Yokohama, Japan}

\begin{document}


\title{Performance and Metacognition Disconnect when Reasoning in Human-AI Interaction
}


\settopmatter{authorsperrow=3}

\author{Daniela Fernandes}
\affiliation{
\institution{Aalto University}
\country{Finland} 
}
\author{Steeven Villa}
\affiliation{
\institution{LMU Munich}
\country{Germany} 
}
\author{Salla Nicholls}
\affiliation{
\institution{Aalto University}
\country{Finland} 
}
\author{Otso Haavisto}
\affiliation{
\institution{Aalto University}
\country{Finland} 
}
\author{Daniel Buschek}
\affiliation{
\institution{University of Bayreuth}
\country{Germany} 
}
\author{Albrecht Schmidt}
\affiliation{
\institution{LMU Munich}
\country{Germany} 
}
\author{Thomas Kosch}
\affiliation{
\institution{HU Berlin}
\country{Germany} 
}
\author{Chenxinran Shen}
\affiliation{
\institution{Independent Researcher}
\country{Vancouver, Canada}
}
\author{Robin Welsch}
\affiliation{
\institution{Aalto University}
\country{Finland} 
}

\makeatletter
\let\@authorsaddresses\@empty 
\makeatother

\renewcommand{\shortauthors}{Trovato et al.}

\begin{abstract}

Optimizing human-AI interaction requires users to reflect on their own performance critically. Our paper examines whether people using AI to complete tasks can accurately monitor how well they perform. In Study 1, participants (N = 246) used AI to solve 20 logical problems from the Law School Admission Test. While their task performance improved by three points compared to a norm population, participants overestimated their performance by four points. Interestingly, higher AI literacy was linked to less accurate self-assessment. Participants with more technical knowledge of AI were more confident but less precise in judging their own performance. Using a computational model, we explored individual differences in metacognitive accuracy and found that the Dunning-Kruger effect, usually observed in this task, ceased to exist with AI. Study 2 (N = 452) replicates these findings. We discuss how AI levels metacognitive performance and consider consequences of performance overestimation for interactive AI systems enhancing cognition.
\end{abstract}


\received{12 September 2024}
\received[revised]{\textit{under revision}}
\received[accepted]{\textit{tbd}}

\begin{teaserfigure}
\centering
  \includegraphics[width=.9\linewidth]{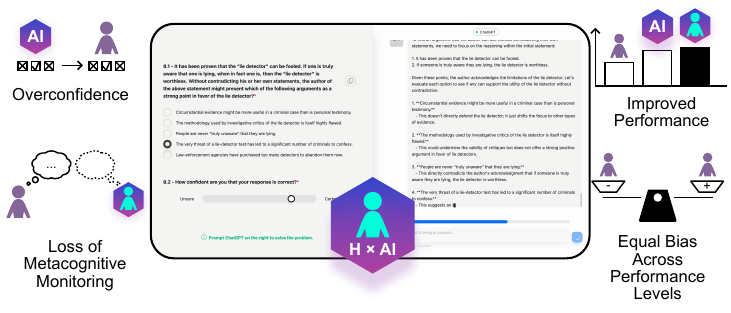}
  \caption{Schematic visualization of our main findings. Participants using AI (ChatGPT-4o) were overconfident and overestimated their performance, although it improved compared to no AI use. The Dunning-Kruger effect was reduced with AI use.}
  \Description[Schematic Figure]{Schematic visualization of our main findings. An interface in the middle with symbols around. Participants using AI (ChatGPT-4o) were overconfident. They overestimated their performance, yet their performance improved. The Dunning-Kruger effect was reduced when assisted by AI to complete the task.}
  \label{fig:teaser}
\end{teaserfigure}
\maketitle

\section{Introduction}



Humans have always used technologies to extend their abilities \cite{clark2010extended,clark2010supersizing,alexandre2024neurohacking}.
Recent advances have aimed to improve human performance and productivity in a range of contexts \cite{zulfikar_memoro_2024, hou_my_2024,wang_human-human_2020, perera_enhancing_2024}. While there is evidence for an improvement in human performance with AI~\cite{bansal_does_2021, steyvers_bayesian_2022, zulfikar_memoro_2024}, these performance gains are limited by effective Human-AI Interaction (HAI). Fundamental biases, such as overtrust and overreliance, impair performance \cite{inkpen_advancing_2023} up to the point that the interaction decreases overall performance as compared to having no AI at all \cite{bastani2024generative,vaccaro2024combinations}. This raises an important question of whether users can and do monitor their AI-augmented performance.

Human metacognition research investigates the ability to monitor, evaluate, and regulate our own cognitive processes~\cite{fleming2024metacognition} and, therefore, has been proposed to be essential in interactive generative AI systems \cite{tankelevitch2023metacognitive}. Psychological research on metacognition has shown that people typically estimate themselves to be better than average \cite{brown1986evaluations}, also called the "better-than-average effect" (see \cite{zell2020better}). 

In the context of AI, people believe AI improves performance \cite{kloft2023ai}, that AI predictions outperform professionals \cite{shekar2024people} and hope AI will improve their lives \cite{cave2019hopes}. 
Research on HAI shows evidence of deficiencies in metacognitive monitoring: users are largely unaware of their performance and their performance improvement with AI. Concretely, when using AI systems, users tend to overestimate their benefits, even when using a sham AI system~\cite{villa2023placebo,kloft2023ai,kosch2023placebo}. 
However, accurate metacognitive monitoring is crucial for optimizing HAI. Inaccurate evaluation of human-AI composite performance \cite{engelbart1962augmenting} can lead to an overreliance on the system, resulting in suboptimal outcomes. From a rational perspective \cite{oulasvirta2022computational}, optimal interaction with AI requires that users possess a clear understanding of their performance to adjust their behavior. 

Similarly, metacognitive judgments exhibit considerable individual variability \cite{ackerman2017meta,toplak2011cognitive}, which can relate to cognitive performance \cite{toplak2011cognitive}. The Dunning-Kruger Effect (DKE) describes a cognitive bias where individuals with lower ability overestimate their competence while those with higher ability underestimate it \cite{kruger1999unskilled}. 
For HAI, a DKE would suggest that low performers may not optimize their interaction with AI due to poor metacognitive monitoring.
However, one could argue that if AI interaction improves overall cognitive performance by augmenting intellect \cite{engelbart1962augmenting}, then metacognitive bias and its link to cognitive performance may disappear. To disambiguate these opposing arguments, we must empirically explore metacognitive monitoring, including metacognitive bias, individual metacognitive accuracy, metacognitive sensitivity, and its relation to performance (DKE) in HAI. We thus conceptually replicate \citet{jansen2021rational} in interaction with AI to explore whether AI impacts self-assessments of performance (\textbf{RQ1}: Does interaction with AI affect metacognitive accuracy?), if it affects the ability to distinguish between correct and incorrect judgments (\textbf{RQ2}: Does interaction with AI affect metacognitive sensitivity?), and if it amplifies or reduces self-assessment bias between low- and high-performing individuals (\textbf{RQ3}: Does interaction with AI affect the DKE pattern?).

We investigate metacognitive monitoring and the opposing perspectives on the DKE by measuring how AI affects users' metacognitive judgments. We designed an experiment where participants used AI to complete logical reasoning tasks from the Law School Admission Test (LSAT). This setting is analogous to those utilized in prior research on metacognitive abilities and the DKE \cite{jansen2021rational}.
By analyzing participants' self-assessments after AI interaction, we can directly measure the AI's influence on participants' metacognition and study its relation to task performance (DKE).




In Study 1, we found that while using AI improves task performance in the LSAT, it also leads to a large overestimation of users' performance (low metacognitive accuracy). Yet, we show using a computational model that the DKE is not only smaller but disappears entirely in our sample while being present in a comparable large-scale sample without AI use \cite{jansen2021rational}. Technological knowledge and critical appraisal of AI, as measured by the "Scale for the assessment of non-experts AI literacy" (SNAIL) \cite{laupichler2023development}, increased confidence but decreased the accuracy of self-assessment. In Study 2 (N = 452), where we incentivize metacognitive monitoring with monetary benefits and collect our own non-AI baseline group, we replicate the pattern of results of Study 1. 

To summarize, although AI has the potential to improve performance and level individual biases in metacognition, it carries the risk of inflated performance self-assessments. Our paper extends our understanding of metacognitive monitoring in HAI by investigating the interplay between metacognition, cognitive performance, and AI literacy. While AI had great benefits for performance and leveled metacognitive deficits, it also brought about the bane of overestimating one's performance. We discuss how to navigate this trade-off and how to improve metacognitive accuracy to empower users to make better decisions when using interactive AI. 
Our contributions and research results are:
\vspace{-.1cm}
\begin{itemize}[itemsep=0pt, parsep=0pt]
    \item Experimentally investigating the impact of AI use on metacognitive monitoring.
    \item Revealing that while AI can improve task performance, it leads to overestimation of performance.
    \item Demonstrating that the DKE is reduced when participants use AI, suggesting that AI can level cognitive and metacognitive deficits.
    \item Highlighting a paradox where higher AI literacy relates to less accurate self-assessment, with participants being more confident yet less precise in their performance evaluations.
    \item Offering design recommendations for interactive AI systems to enhance metacognitive monitoring by empowering users to critically reflect on their performance.
\end{itemize}



\section{Related Work}

In the following section, we focus on the role of metacognitive processes in human cognition and human-AI interaction. We introduce central concepts of human metacognition and examine how self-monitoring and AI assistance can impact cognitive performance and user decision-making.

\subsection{Human Metacognition}
Metacognitive processes (e.g., monitoring or evaluating one's cognition) are crucial for problem-solving, learning, and behavior optimization \cite{fleming2024metacognition}. Metacognitive judgments primarily involve accuracy and sensitivity: accuracy measures how closely self-evaluations match actual performance, enabling better decisions when recognizing limitations \cite{fleming2024metacognition,colombatto2023illusions}, while sensitivity assesses the ability to distinguish between correct and incorrect judgments, often via confidence ratings \cite{fleming2024metacognition}.

Metacognitive accuracy is shaped by bias (systematic over/underestimation) and noise (random fluctuations) \cite{fiedler2019metacognition}, where bias skews evaluations and noise reduces sensitivity \cite{fleming2024metacognition}.
The Dunning–Kruger effect (DKE) links metacognitive accuracy with skill level: low-performing individuals overestimate their performance, while high performing individuals underestimate theirs \cite{kruger1999unskilled}. Although some argue the DKE may be a statistical artefact \cite{gignac2020dunning,gignac2024rethinking}, studies like \citet{jansen2021rational} and \citet{ehrlinger2008unskilled} replicated these findings with large samples, confirming the DKE in verbal and logical reasoning tasks.

Although metacognition and the DKE are frequently examined in educational contexts (e.g., see \citet{yang2024competence} in math education and \citet{mahdavi2014overview} for an overview), our approach to studying the DKE shifts the focus to a task-specific context. Following \citet{dunning2011dunning}, we replicate the method of \citet{jansen2021rational} by concentrating on task performance and ratings of absolute performance estimates after task completion. 

\textbf{In sum, metacognitive abilities are central to optimizing cognition and explain performance differences underlying the DKE.}

\subsection{Metacognition in Human-AI Interaction}
Augmenting human intellect has been central to HCI, as highlighted by Engelbart and others \cite{engelbart1962augmenting,schmidt2017augmenting} and metacognition has been studied, albeit seldom quantitatively, at CHI \cite{Tombaugh1985,tankelevitch2023metacognitive,Quinn1986,chan2014social,corbett2001locus,zhang2022,foster2013competition,seitlinger2012implicit,gooch2016using,loksa2016programming,lu2021human}. 

Augmenting human cognitive processes through AI has recently received attention in CHI. Augmentation has been explored in writing \cite{reza2024abscribe,goldi2024intelligent,li2024thevalue,lee2024designspace}, reading comprehension \cite{xiao2023inform}, reasoning \cite{danry2023dontjust} and solving math arithmetic tasks \cite{xu2023augmenting}. In addition, research has addressed the psychological challenges in augmenting humans with AI, such as appropriate reliance on AI \cite{he2023knowing}, trust in AI and its impact on performance \cite{ahn2024impact}, and inefficiencies in task delegation to AI systems \cite{pinski2023aiknowledge}. However, there is a notable lack of consideration for metacognition \cite{tankelevitch2023metacognitive}, which refers to how people monitor their own performance when augmented with AI and what brings optimal performance in human-AI interaction.

A recent survey by \citet{vaccaro2024combinations} distinguishes between human-AI synergy -- where combined performance surpasses either humans or AI alone -- and human-AI augmentation, where humans aided by AI do better than unassisted humans. They found that when humans already outperform AI, adding AI improves the team's overall performance. However, as AI becomes more powerful, the average performance of these teams declines. Thus, a central challenge in human-AI interaction is achieving synergy when AI models surpass human capabilities.

For LLM's, studies have measured AI's benefits to human cognitive performance, such as \citet{Shakked2023} finding a 40\% increase in students' essay-writing speed with AI, and \citet{bastani2024generative} reporting a 48\% improvement in math performance with ChatGPT. However, students using ChatGPT still performed 17\% worse in exams compared to a control group \cite{bastani2024generative}, indicating that AI can sometimes hinder skill development \cite{bastani2024generative,rafner2022deskilling,Kobiella2024}.

These issues likely stem from suboptimal interfaces that fail to support metacognition \cite{tankelevitch2023metacognitive,kosch2023placebo,kloft2023ai}. Research shows that users often overestimate their AI-assisted performance and struggle to monitor or plan interactions effectively \cite{kosch2023placebo,kloft2023ai,villa2023placebo,Bosch2024}. For instance, \citet{johnny2023} found that users have difficulty crafting effective prompts, while \citet{dang2023choice} noted challenges in switching between tasks and writing prompts. Furthermore, explanations from AI systems are often uninformative, ignored, or lead to cognitive biases themselves \cite{eiband2019impact,wang2021explanations,vasconcelos2023explanations,bertrand2022cognitive}. AI literacy, which involves understanding AI concepts and evaluating outputs critically, is also essential for effective interaction \cite{laupichler2023development}. However, its influence on metacognitive judgements in AI-assisted decision-making and interaction optimization is unclear. 
Therefore, a key limitation in human-AI interaction is primarily metacognitive, involving difficulties in planning, monitoring, and evaluating interactions. \textbf{Theoretically, AI imposes new metacognitive demands on users \cite{tankelevitch2023metacognitive}, requiring enhanced monitoring and control. However, empirical research on metacognitive monitoring is scarce.}

\section{Study 1: Metacognition in Human-AI interaction}
We have conducted two studies. One compares a group of participants using AI to \citet{jansen2021rational} data on the LSAT (that have not used AI); the other study, Study 2, replicates and extends Study 1. 

\subsection{Method}

In the following, we motivate and document our methodological choices when conducting Study 1. The research software, the analysis with all associated measures and data, can be found at [blinded for review].  Note that both the data and the material of \citet{jansen2021rational} are openly available under \href{}{https://osf.io/er9ms/}, which allowed us to closely follow their task environment and sample characteristics for the purpose of Study 1.  All data collected for the purpose of our paper and analysis scripts can be found at \osfrepository.

\subsubsection{Participants}

To explore individual differences in cognitive and metacognitive performance, we recruited a larger sample than typical DKE studies, allowing us to detect differences in metacognitive accuracy across high and low performers \cite{dunning2011dunning,gignac2024defining}. 
We powered for the smallest effect of interest, which is the DKE. For power analysis, we used bootstrapped samples of \citet{jansen2021rational} with sample sizes ranging from 80 to 400 to assess the ability to detect the DKE through \textit{t}-tests across quartiles. We computed the proportion of \textit{p}-values < .05 to determine the optimal sample size for sufficient statistical power (80\%). With this, we found that a sample size of 250-300 participants is optimal for reliably detecting differences between the upper and lower quartiles in metacognitive accuracy.. 

We recruited 274 English-fluent participants located in the USA through Prolific. We included an attention check, requiring participants to read a short description of the study and task. They then answered two multiple-choice questions, one about the topic (logical reasoning) and another regarding which option to choose when solving the problems (the best one).
We excluded thirteen (13) participants due to failing the attention check, as well as two (2) due to erroneous responses (e.g., exceeding the number of possible correct answers in estimating performance) and thirteen (13) due to low completion times.

We further analyzed data from 246 participants (identified as female 114; identified as male 130, identified as non-binary 2; Age: $M$ = 39.85, $SD$ = 14.53). When asked to estimate their English fluency, 
218 participants reported themselves as native English speakers, 25 as fully fluent, two as conversationally fluent and one as understanding basic English. No participants preferred not to disclose their language proficiency. 14 participants in our sample reported their highest educational degree to be a doctoral degree, 58 a higher tertiary education degree (Master's level), 95 a lower tertiary education degree (Bachelor's level), 52 an upper secondary school / high school and 27 a vocational college degree. 
13 participants had taken the LSAT before; their performance was slightly lower, $M$ = 12.38, $SD$ = 3.22, compared to those who have not taken it, $M$ = 13.01, $SD$ = 2.85, thus they were not excluded from the sample. We collected informed consent from each participant before the study in accordance with the Declaration of Helsinki guidelines \cite{world2013world}. Each participant was compensated 6.5 pounds per hour.  In accordance with the [Blinded] national guidelines (the [Blinded] National Board on Research Integrity), this study did not require ethics approval as it involved minimal risk to participants, with no intervention beyond standard practice and no collection of sensitive personal data. 



\subsubsection{Materials}

Participants’ logical reasoning ability was measured with the 20 multiple-choice logical reasoning problems used by \citet{jansen2021rational} to approximate the LSAT a widely recognized, real-world assessment used in high-stakes decision-making, such as law school admissions \cite{shultz2011predicting,wainer1995precision}. It also serves as a benchmark in machine learning research, making it ideal for comparing AI-assisted performance \cite{katz2024gpt}. 
 An example LSAT question provided to participants was:
"it has been proven that the lie detector can be fooled. If one is truly aware that one is lying, when in fact one is, then the lie detector is worthless.
The author of this argument implies that: (1) The lie detector is sometimes worthless.
(2) The lie detector is a useless device. (3) No one can fool the lie detector all of the time. (4) A good liar can fool the device. (5) A lie detector is often inaccurate.".

Using the same items as \citet{jansen2021rational} enabled us to compare our results to a representative sample of participants who did not use AI in the task and replicate the results of the original study by \citet{kruger1999unskilled}. 

In addition to participants' actual logical reasoning performance with AI use, perceived performance with and without AI, as well as the AI system’s performance on its own, was measured using the items presented in Table \ref{tab:descriptives}. Lastly, we measured participants' AI literacy using the SNAIL \cite{laupichler2023development} at the end of the study. The scale features 31 items to assess participants' technical understanding, critical appraisal and practical application of AI systems. The scores can be found at the end of Table \ref{tab:descriptives}.

\subsubsection{Task}\label{method:task}

Participants completed 20 LSAT logical reasoning items in a randomized order. Each problem was displayed on the left-hand side of the screen, while a ChatGPT interface was presented on the right. 
Participants were required to interact with ChatGPT for assistance, ensuring at least one prompt per problem, before submitting their final answers and rating their confidence in their response ("How confident are you that your response is correct?"; from "unsure" to "certain" on a 100-step slider). Unlimited text chat interaction was enabled during the task, allowing participants to engage with ChatGPT as much as they deemed necessary. The LSAT problems were intended to assess logical reasoning abilities and did not require any prior knowledge of law to solve.



\subsubsection{Apparatus}
\begin{figure}
    \centering
    
    \centering
    \includegraphics[width=1\linewidth]{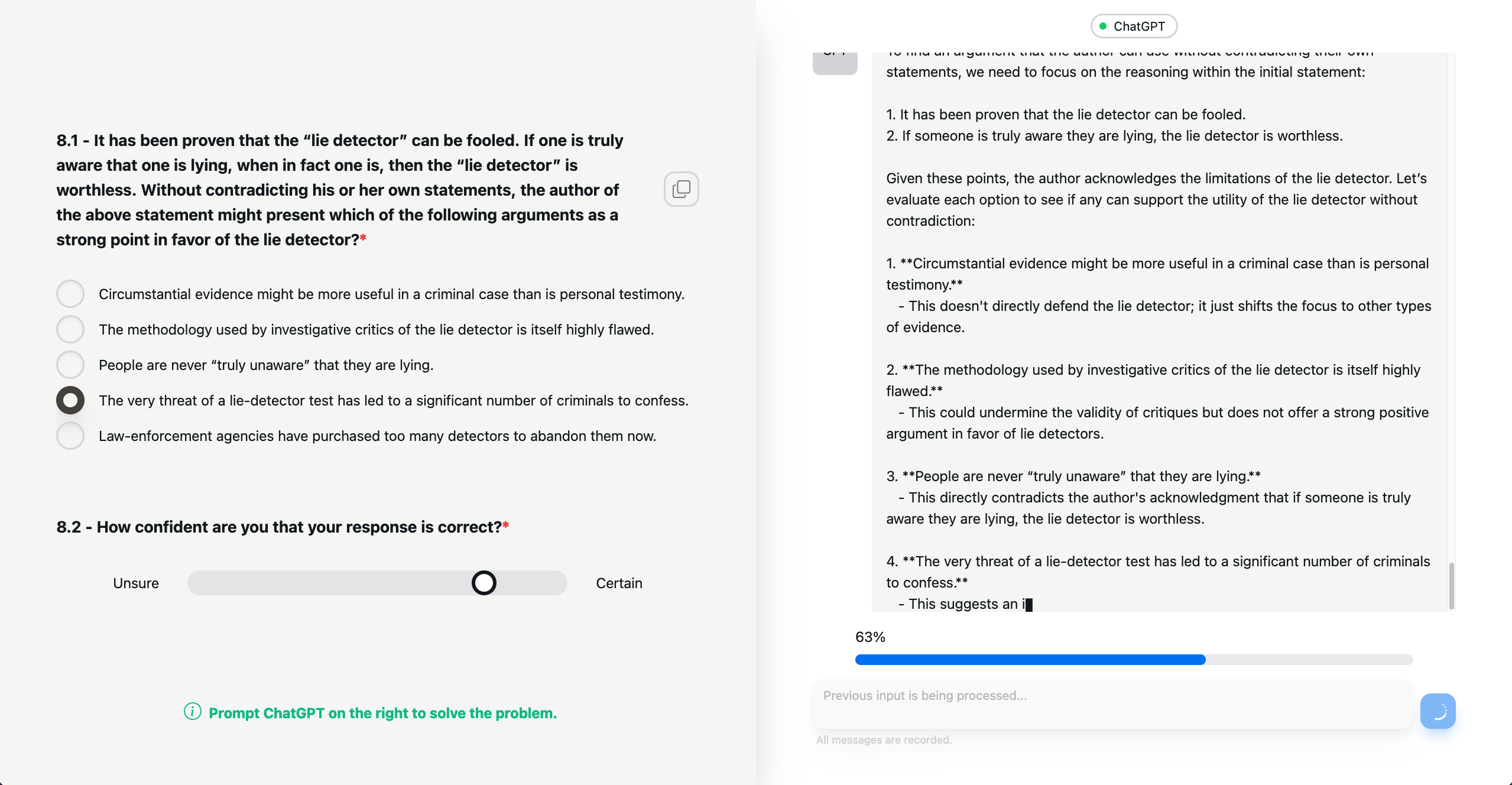}
    \caption{Our online study application featured a horizontally split interface, with survey items and logical reasoning problems presented on the left and a ChatGPT interface on the right.}\label{fig:study-ui}
    \Description{Our online study application featured a horizontally split interface, with survey items and logical reasoning problems presented on the left and a ChatGPT interface on the right.}
\end{figure}

We inspected the software and flow of \citet{jansen2021rational} and carefully replicated it, integrating a side-by-side view of ChatGPT and survey interface (Figure \ref{fig:study-ui}).
We used ChatGPT-4o (\verb|gpt-4o-2024-05-13|) due to its widespread use in enhancing cognitive task performance \cite{bastani2024generative,draxlerghost24,draxler2023gender}. A custom interface (\autoref{fig:study-ui}) was built to log user interactions, enabling qualitative chat analysis.  The application automatically collected survey responses and chat logs and recorded them at the end of the study. We included a button to copy each problem and its answer options to the clipboard so they could be easily pasted into the chat.


\subsubsection{Procedure}
Upon entering the study on Prolific, participants were redirected to our application. After consenting to participate, we asked our participants to complete the user expectations questionnaire. Then, they were briefly introduced to the task and allowed to test the chat interaction. Afterwards, participants engaged in a task to assess their logical reasoning skills by solving a series of LSAT problems. Before submitting their final answers, they were asked to interact with ChatGPT if assigned to the AI group, ensuring they provided at least one prompt per problem. After completing each question, participants rated their confidence in their response using a 100-step slider ranging from ``unsure'' to ``certain.'' In the AI group, participants were permitted unlimited text-based interaction with ChatGPT, allowing participants to seek as much assistance as they felt necessary during the problem-solving process.  After solving the problems in a randomized order (\ref{method:task}), participants were again asked to complete the expectations questionnaire in the past tense. They also responded to the SNAIL questionnaire \cite{laupichler2023development} and filled in their demographic information, including age, gender, occupation, education level, English proficiency, and whether they had taken the LSAT before. Study 1 took, on average, 42 minutes to complete.


\subsubsection{Analysis}

We analyzed the data in five steps. First, we compared our sample performance to that of \citet{jansen2021rational} and to the performance of AI alone. This allowed us to analyze where AI augments human performance (i.e. human-AI interaction outperforms a no-AI group) and human-AI synergy (i.e. human-AI interaction outperforms AI). We analyze both average performance and its distribution. Second, we analyzed metacognitive accuracy on a task level -- the difference between objective and estimated performance\footnote{In line with \citet{ehrlinger2008unskilled}, we focus on numeric estimates of performance after the task, not relative performance in comparison to others.} comparing human-AI interaction to a no-AI group -- and trial-level confidence ratings to assess metacognitive sensitivity. With this, we can analyze metacognitive performance of users when interacting with AI globally (accuracy) after the tasks and locally (sensitivity) after each decision. Note that for \citet{jansen2021rational}, the question was ("How many of the 20 logical reasoning problems do you think you solved correctly?") while in our sample using AI, we asked: "Using the AI, how many of the 20 logical reasoning problems do you think you solved correctly?". 
Next, we correlated metacognitive performance metrics with performance and AI literacy measures to explore what predicts low metacognitive performance in human-AI interaction (for a similar analysis approach see \citet{mcintosh2019wise}). Fourth, we used a computational model of performance and performance assessments to compare our sample and \citet{jansen2021rational} to estimate how AI affects the DKE. Lastly, we qualitatively analyzed participant strategies and how they conceptualize the human-AI relation\footnote{Given the large samples in our study and the ceiling effects encountered \autoref{fig:corplot}, we do not test for the normality of residuals of our variables. Instead, we model the data using a final Bayesian computational framework, which allows for more flexible assumptions and can account for ceiling effects in our main analysis. This approach provides more robust estimates by incorporating uncertainty in a probabilistic manner rather than relying on strict parametric assumptions.}. We use frequentist statistics ($\alpha $= 5\%) for simple statistical tests (e.g., paired and unpaired \textit{t}-test and Pearson correlation) in the first four analyses and Bayesian statistical modelling for the computational model. Note that we focus only on these analyses in the exploration of our data for the current paper; however, we encourage the re-analysis and further exploration in the openly available dataset \osfrepository.

\subsection{Results}




\sisetup{parse-numbers = false}
\begin{table}
\centering
\caption{Descriptive Statistics for all subjective variables for the full sample and split by performance quartile (Q). }\label{tab:descriptives}
\Description{Descriptive statistics for various subjective variables, divided by quartiles (Q1 to Q4) of participant performance. It includes data on estimated and actual performance, AI system evaluations, and task difficulty. The full sample size is 246, and values like average performance, confidence, and AI effectiveness estimates are provided. Metrics related to AI literacy (SNAIL: Technical Understanding, Critical Appraisal, and Practical Application) are also shown for each quartile.}
\centering
\resizebox{\ifdim\width>\linewidth\linewidth\else\width\fi}{!}{
\fontsize{12}{14}\selectfont
\begin{tabular}[t]{p{15cm}S[table-format=2.2(2.2)]|S[table-format=2.2(2.2)]S[table-format=2.2(2.2)]S[table-format=2.2(2.2)]S[table-format=2.2(2.2)]}

 & \text{Full sample} & Q1 & Q2 & Q3 & Q4\\
\toprule
\textit{n} & 246 &110 & 59 & 55 &22 \\
Performance & 12.98 (2.88) & 10.82 (3.07) & 14 (-) & 15 (-) & 16 (-)\\
Estimate & 16.50 (3.71) & 15.34 (4.31) & 17.39 (2.93) & 17.40 (3.07) & 17.68 (2.17)\\
\hdashline
Compared to other participants in this study, how would you rate your general logical reasoning ability when using the help of AI? (\% rank) & 68.08 (19.3) & 66.02 (20.51) & 70.71 (18.35) & 68.22 (17.87) & 71.0 (19)\\  &&&&&\\
Using the AI, how many of the 20 logical reasoning problems do you think you will solve correctly? & 15.96 (3.63) & 15.49 (3.83) & 16.17 (3.6) & 16.33 (3.68) & 16.82 (2.13)\\  &&&&&\\
Without AI use, how many of the 20 logical reasoning problems do you think you would solve correctly? & 11.64(4.53) & 11.25(4.95) & 11.36(4.19) & 12.24(4.31) & 12.91(3.58)\\  &&&&&\\
Compared to other AI systems, how would you estimate the AI system's logical reasoning ability? (\% rank) & 70.02 (17.91) & 69.04 (18.04) & 69.44 (18.01) & 70.0 (18.27) & 76.59 (15.79)\\  &&&&&\\
On its own, how many of the 20 logical reasoning problems do you think the AI would solve correctly? &16.54 (7.75) & 15.75 (4.53) & 16.47 (3.74) & 18.31 (14.42) & 16.32 (3.26)\\  &&&&&\\
Compared to other participants in this study, how well do you think you will do? (\% rank) & 66.95 (19.84) & 64.0 (21.93) & 69.31 (17.49) & 68.8 (17.54) & 70.77 (19.3)\\ 
&&&&&\\
How difficult is solving logical reasoning problems for you? & 5.38 (2.01) & 5.43 (2.05) & 5.88 (2.03) & 4.93 (1.86) & 4.91 (1.93)\\  &&&&&\\
How difficult is solving logical reasoning problems for the average participant? & 6.09 (1.56) & 6.14 (1.62) & 6.34 (1.59) & 5.93 (1.53) & 5.64 (1.14)\\  &&&&&\\
Compared to other participants in this study, how would you rate your general logical reasoning ability when using the help of AI? (\% rank) & 69.83 (21.86) & 66.86 (24.44) & 72.9 (19) & 72.2 (20.18) & 70.45 (18.5) \\   &&&&&\\
\hdashline 
Using the AI, how many of the 20 logical reasoning problems do you think you solved correctly? & 16.5 (3.72) & 15.35 (4.31) & 17.39 (2.93) & 17.4 (3.07) & 17.68 (2.17)\\  &&&&&\\
Without AI use, how many of the 20 logical reasoning problems do you think you would have solved correctly? & 11.61 (4.52) & 11.21 (4.6) & 11.78 (4.47) & 12.00 (4.94) & 12.18 (3.03)\\  &&&&&\\
Compared to other AI systems, how would you estimate the AI system's logical reasoning ability? (\% rank) & 76.29 (18.42) & 74.04 (19.51) & 78.41 (17.03) & 77.2 (19.57) & 79.64 (11.92)\\  &&&&&\\
On its own, how many of the 20 logical reasoning problems do you think the AI would have solved correctly? & 17.74 (8.65) & 17.25 (10.4) & 17.68 (2.84) & 18.69 (10.45) & 18.0 (2.16)\\  &&&&&\\
Compared to other participants in this study, how well do you think you performed? (\% rank) & 68.63 (21.39) & 65.11 (23.81) & 71.14 (19.1) & 72.13 (19.21) & 70.82 (17.9)\\  &&&&&\\
How difficult was solving these logical reasoning problems for you? & 5.67 (2.33) & 5.85 (2.35) & 5.64 (2.24) & 5.47 (2.39) & 5.32 (2.36)\\  &&&&&\\
How difficult was solving these logical reasoning problems for the average participant? & 6.11 (2.09) & 6.3 (2.14) & 6.07 (1.95) & 5.82 (2.15) & 6.0 (2.09)\\  &&&&&\\ \hdashline 
SNAIL: Technical Understanding & 3.83 (1.6) & 3.75 (1.57) & 3.99 (1.62) & 3.77 (1.66) & 3.94 (1.55)\\  &&&&&\\
SNAIL: Critical Appraisal & 5.03 (1.28) & 5.02 (1.29) & 5.05 (1.25) & 5.01 (1.33) & 5.05 (1.29)\\  &&&&&\\
SNAIL: Practical Application & 5.02(1.31) & 5.0(1.39) & 4.99(1.27) & 5.02(1.32) & 5.17(1.07) \\
\bottomrule

\end{tabular}}

\raggedright Note: M (SD) and the sample size (n). Scale for the assessment of non-experts’ AI literacy (SNAIL). According to task context, rank \% instructions are scaled as follows: marking 90\% means you perform better than 90\% of participants, marking 10\% means you perform better than only 10\% of participants, and marking 50\% means you will perform better than half of the participants.- indicates no variation, e.g., when all participants in a quantile had the same value. 

\end{table}


\subsubsection{Human-AI Composite Performance}
\begin{figure}
    \centering
    \includegraphics[width=\linewidth]{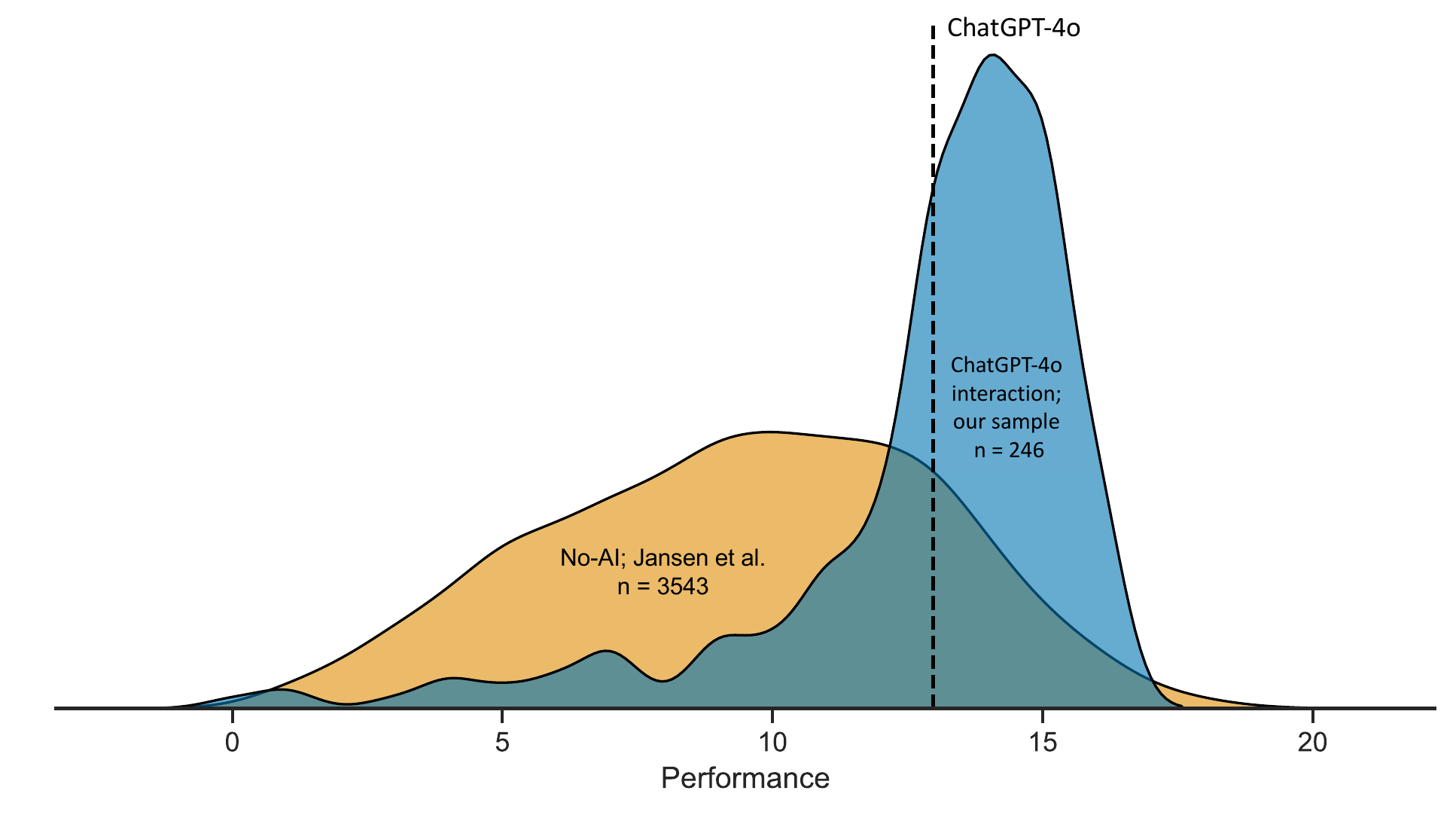}
    \caption{Comparison of performance scores between participants interacting with ChatGPT and a dataset without AI \cite{jansen2021rational}. The blue density curve represents the distribution of performance scores in a sample of 246 participants using ChatGPT, showing a peak in performance at around 14 points. The yellow density curve corresponds to a larger sample (n = 3543) from the \citet{jansen2021rational} dataset, with lower overall performance scores. The vertical dashed line indicates the mean performance score in the ChatGPT simulation.}
    \label{fig:synergyperformance}
    \Description{The graph compares performance scores between participants using ChatGPT and those without AI assistance from the Jansen et al. dataset using density plots. The ChatGPT group shows higher performance, peaking around 14 points, while the no-AI group has lower overall scores. A dashed line marks the mean performance in the ChatGPT simulation. One can clearly see that performance is increased for ChatGPT interaction and that this distribution is much more narrow than Jansen's data.}
\end{figure}

To see whether there is a synergy effect of using ChatGPT in the LSAT (i.e. Human-AI performance > AI performance) we compared the average ChatGPT performance (100 runs at $M$ = 13.65;) to our user's average performance ($M$ = 12.98, $SD$ = 2.88). We find a significant difference, with participants performing slightly worse than ChatGPT alone $t(245) = -3.66, p < .001, d = -0.23$ in the task. 
Next, we compared our sample's performance to \citet{jansen2021rational} representative sample of 3543 participants, who completed the same task without any assistance ($M$ = 9.45, $SD$ = 3.59). We find that in our sample, participants performed significantly better with ChatGPT assistance $t(245) = 19.23, p < .001, d = 1.23$ as compared to the \citet{jansen2021rational} sample. Therefore, while, on average, there is no human-AI synergy, we do find that ChatGPT use can augment human performance for solving the LSAT logical problems (i.e. human-AI performance > human performance).

Looking at individual performance, we can find indications of human-AI synergy.  The difference between our sample's and ChatGPT's performance is rather small, at less than one point. 55.28\% (136 of 246) of our participants performed better than ChatGPT. However, 89.43\% (220 of 246) in our sample performed better than the average score of the \citet{jansen2021rational} sample (see also \autoref{fig:synergyperformance}). 

Therefore, while overall performance increased with the use of ChatGPT augmenting the human ability to solve LSAT problems, on average, we do not find a human-AI synergy. The composite performance of ChatGPT and the participant overtook the performance of ChatGPT alone for only slightly more than half of the participants in our sample. With our prototype's ability to enhance performance established in human augmentation but not synergy, we can now focus on investigating metacognitive abilities. 

\subsubsection{Metacognitive accuracy and sensitivity}

Participants were inaccurate in assessing their performance after task completion, as indicated in the item ``Using the AI, how many of the 20 logical reasoning problems do you think you solved correctly?'', see also \autoref{tab:descriptives}. On average, they estimated solving about 17 out of 20 items  ($M$ = 16.50, $SD$ = 3.72). This overestimation of about 4 points could be distinguished from 0, $t(256) = 14.98, p < .001, d = 0.93$. 

\begin{figure}
    \centering
    
    \includegraphics[width=1\linewidth]{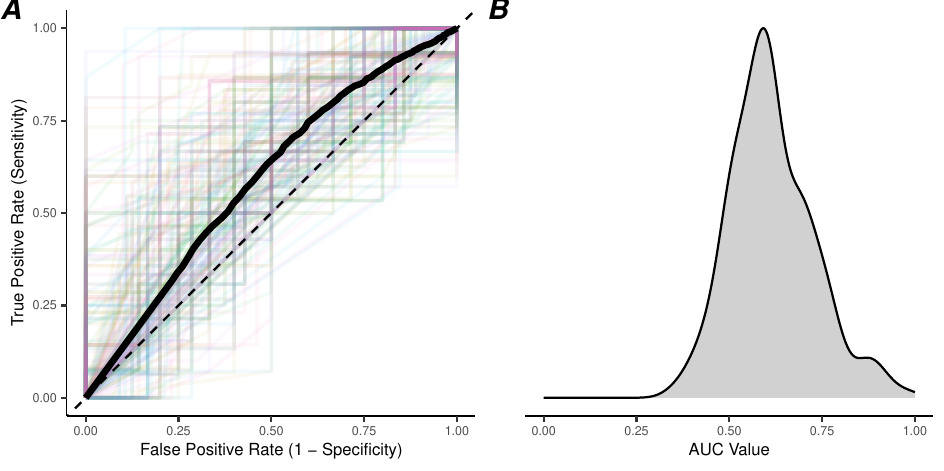}
    \caption{(A) Receiver Operating Characteristic (ROC) curves showing the relationship between True Positive Rate (Sensitivity) and False Positive Rate (1 – Specificity) for participants (color) and pooled across participants (black). The dashed diagonal line indicates the line of no discrimination (random guessing). For each participant, we generated a ROC curve, colored lines, as well as one from pooled responses (black line), which illustrates the trade-off between the true positive rate (i.e., the proportion of correct judgments identified as correct) and the false positive rate (i.e., the proportion of incorrect judgments identified as correct) across various confidence thresholds. 
    (B) Distribution of Area Under the Curve (AUC) values across all participants, with a peak around 0.6, suggesting variability in metacognitive sensitivity, with most participants performing above chance level (AUC = .5).}
    \label{fig:ROCAUC}
    \Description{Two plots related to metacognitive sensitivity. Plot A displays Receiver Operating Characteristic (ROC) curves for participants, illustrating the relationship between True Positive Rate (Sensitivity) and False Positive Rate. With most participants performing better than chance. The dashed diagonal line represents random guessing. For each participant, we generated a ROC curve, as well as one from pooled responses, which illustrates the trade-off between the true positive rate. Plot B shows the distribution of Area Under the Curve (AUC) values across all participants, with a peak around 0.6, indicating variability in metacognitive sensitivity. Most participants perform slightly above chance level (AUC = 0.5).}
\end{figure}

To see whether participants track information in each trial, we turn to metacognitive sensitivity that we estimate from confidence ratings when making the decision. 
Mean confidence (rated on a scale from 0-100) for correct answers was 82.49 (14.24) and for incorrect answers 77.00 (16.52), \textit{t}(244)~=~8.21, \textit{p}~<~.001, \textit{d}~=~0.52. To evaluate sensitivity more granularly, we conducted a Receiver Operating Characteristic (ROC) analysis. ROC analysis is a technique used to assess the performance of a judgment by plotting the true positive rate against the false positive rate across different thresholds.
Here, we applied ROC analysis to understand how well participants' confidence scores predicted whether their responses were correct (see \autoref{fig:ROCAUC}A). 

By using ROC analysis, we obtain a metric, the area under the curve (AUC) that quantifies how effectively a participant’s confidence ratings differentiate between correct and incorrect responses. An AUC value of 0.5 indicates no better-than-chance discrimination, while higher values reflect greater sensitivity—meaning the participant’s confidence reliably tracks correctness. Thus, the ROC analysis provides a nuanced individual, trial-level measure of metacognitive accuracy beyond simple average confidence or aggregate performance estimates. 

The mean AUC was .62 ($SD$ = 11.2) which could be distinguished from 0.5 ($t(244) = 16.02, p < .001, d = 1.02$). While most participants' (210 out of 246; 85.37\%) metacognitive AUC values are above .50 (random guessing). This means that confidence scores indicate participants' metacognitive sensitivity on a trial level in human-AI interaction. Note that for the remaining 36 participants' confidence ratings were not able to distinguish between correct or incorrect trials (for the distribution of AUC values, refer to \autoref{fig:ROCAUC}B). For these participants, confidence judgments were effectively random or worse than random chance, indicating that they tended to be as confident or even more confident about incorrect responses than correct ones. This pattern suggests a miscalibration in their metacognitive judgments, where confidence fails to serve as a reliable indicator of actual performance. Thus, our sample exhibits very low metacognitive sensitivity and, in consequence, low metacognitive monitoring.

\subsubsection{Correlation of metacognitive ability, performance, and AI literacy (SNAIL)}

A number of significant relationships were found when correlating several metacognitive indices with LSAT performance and AI literacy, see \autoref{tab:Correlations}. We observed a positive relation between performance and participants’ average confidence estimates. Participants who perform well are also more confident on average. However, those who are, on average, more confident also overestimate their performance due to increased metacognitive bias. This is probably due to the relationship between SNAIL factors and performance estimates, where participants who express more technical knowledge and more critical appraisal also estimate their performance to be relatively higher. However, those with high technical understanding are also less accurate in their metacognitive judgements. All SNAIL factors correlated positively to average confidence. Note that these correlations are rather small and should thus be interpreted with caution. Metacognitive sensitivity (AUC and $\Delta conf$) was not related to AI literacy, performance or metacognitive accuracy.


\begin{table}
\caption{Correlation Table of Metacognitive measures and AI literacy as measured by the SNAIL}
\label{tab:Correlations}

\centering
\begin{tabular}[t]{llllllllll}
\toprule
  & $\Delta EP$ & Estimate & Performance & $\Delta conf$ & $\mu conf$ & AUC & SNAIL TU & SNAIL CA & SNAIL PA\\
\midrule
$\Delta EP$ &  &  &  &  &  &  &  &  & \\
Estimate & 0.72*** &  &  &  &  &  &  &  & \\
Performance & -0.43*** & 0.32*** &  &  &  &  &  &  & \\
$\Delta conf$ & -0.04& -0.03& 0.01&  &  &  &  &  & \\
$\mu conf$ & 0.24*** & 0.46*** & 0.27*** & -0.10&  &  &  &  & \\
AUC & 0.03 & 0.05 & 0.03 & 0.59***& -0.08 &  &  &  & \\
SNAIL TU & 0.21** & 0.17** & -0.06 & -0.12& 0.13* & -0.10 &  &  & \\
SNAIL CA & 0.10 & 0.14* & 0.06 & 0.04& 0.25*** & 0.03 & 0.49*** &  & \\
SNAIL PA & 0.05 & 0.10 & 0.06 & 0.01& 0.24*** & 0.05 & 0.57*** & 0.81*** & \\
\bottomrule
\end{tabular}
\noindent \raggedright \textbf{Note}. $df$ = 243, * indicates $p$ < .05,** indicates $p$ < .01 and *** indicates $p$ < .001. $\Delta EP$ represents the difference between performance and estimated performance (metacognitive accuracy). Performance refers to the achieved task performance. $\Delta conf$ is the difference between predicted and actual confidence, while $\mu conf$ is the mean confidence (average confidence ratings). AUC refers to Area Under the Curve, with a higher AUC value indicating a more reliable confidence score in reflecting participants' correctness; SNAIL TU stands for the Technical Understanding score, SNAIL CA represents the Critical Appraisal score, and SNAIL PA is the Practical Application score.
\Description{Correlations between metacognitive measures and AI literacy (SNAIL). Key variables include the difference between estimated and actual performance, confidence measures, Area Under the Curve, and three SNAIL components: Technical Understanding score (SNAIL TU), Critical Appraisal score (SNAIL CA), and Practical Application score (SNAIL PA)}
\end{table}

\subsubsection{AI use cancels the Dunning-Kruger effect}

\begin{figure}[!h]
    \centering
    \begin{subfigure}[b]{0.49\linewidth}
        \centering
        \includegraphics[width=\linewidth]{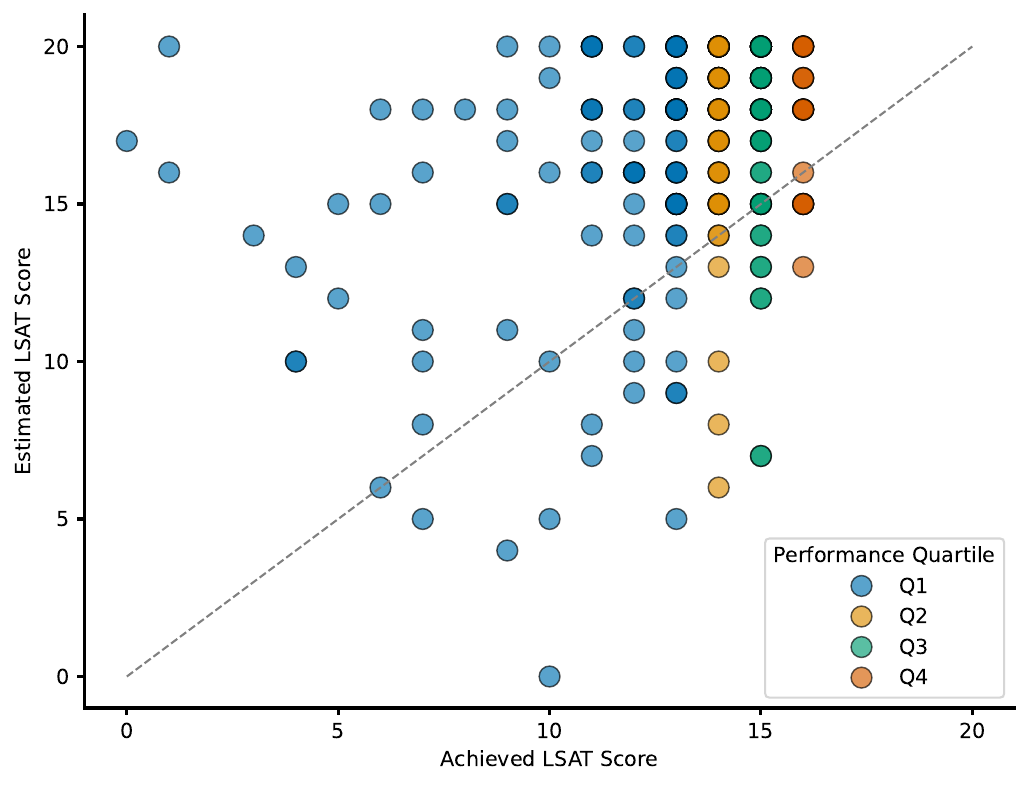}
        \caption{Scatter-plot of Estimated LSAT Scores as a function of Achieved LSAT Scores for the AI group in Study 1.}
        \label{fig:corplot}
        \Description{}
    \end{subfigure}
    \hfill
    \begin{subfigure}[b]{0.49\linewidth}
        \centering
        \includegraphics[width=\linewidth]{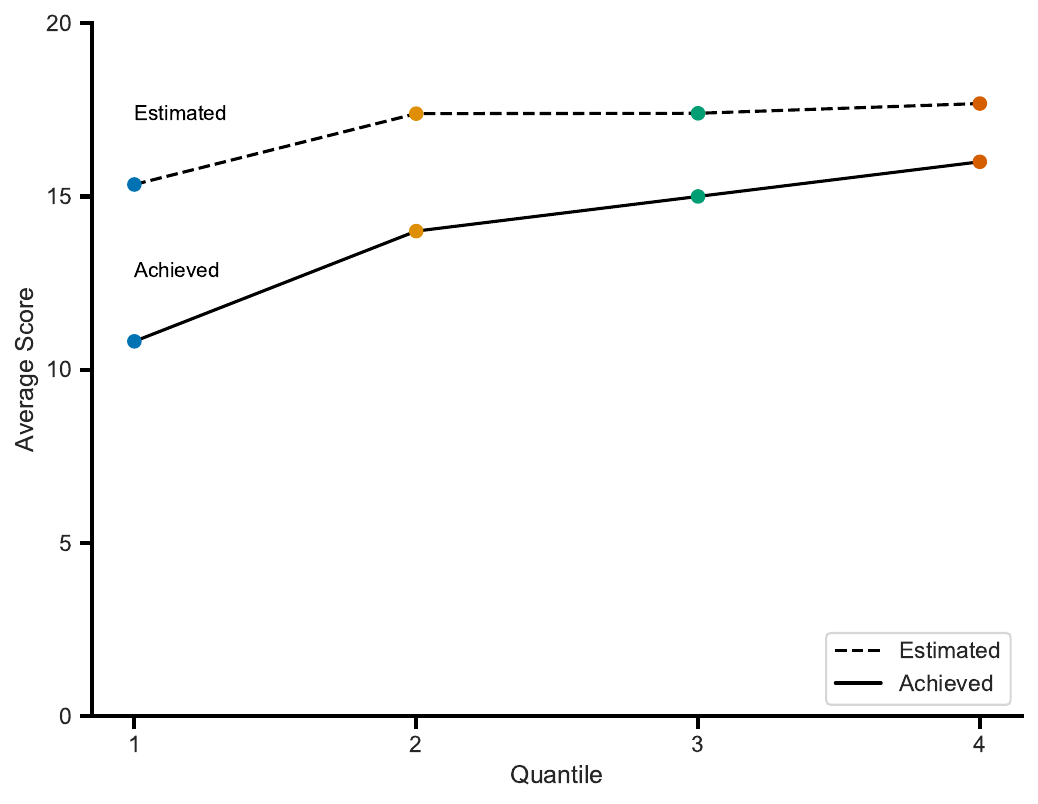}
        \caption{Classical Dunning-Kruger Quartile-plot. Comparison of Mean Estimated and Mean Achieved Scores Across Quartiles for the AI group in Study 1.}
        \label{fig:dkplot}
        \Description{}
    \end{subfigure}
    \caption{Correlation of estimated and achieved LSAT score from different perspectives, individual \autoref{fig:corplot} vs. quartile-level \autoref{fig:dkplot} for the AI group in Study 1.}
    \Description{Plot (a): scatter plot of estimated LSAT scores as a function of achieved LSAT scores across quartiles (Q1 to Q4). While some participants were accurate in their estimates, some participants were very wrong in estimating their performance accurately. Plot (b): quartile-level Dunning-Kruger plot, showing that estimated scores are consistently higher than achieved scores across all quartiles with a spread for low performers and closer match of estimates and performance in high performers.}
    \label{fig:combined}
    
\end{figure}

The correlation between estimated performance and actual performance is small to medium-sized (see \autoref{tab:Correlations}, and for visual representation \autoref{fig:corplot}). While some participants were very accurate in estimating their performance, some participants were considerably off in their estimates (\autoref{fig:corplot}). This suggests the possibility of a DKE-like pattern, where ability in a task is related to the metacognitive ability to judge one's task performance. For the classical quantile plot, refer to \autoref{fig:dkplot}\footnote{Note, however, that this plot can be misleading \cite{gignac2020dunning}}.

We calculated the difference between quantiles to test whether metacognitive accuracy was worse in the low-scoring quantile than in the high-scoring quantile. Both quantile's metacognitive accuracy differed from 0, (Q1: $t(109) = -10.15, p < .001, d = -0.97$, Q4: $t(21) = -3.64, p = .002, d = -0.78$), probably due to the overall bias. However, we found that the difference for Q1 is larger than for Q4, $t(130) = 2.79, p = .006, d = 0.49$ (see also \autoref{fig:dkplot} and \autoref{tab:descriptives}). Note that this pattern of effect could be driven by metacognitive bias alone. To establish a DKE, we must first quantify the metacognitive noise in our sample. To do so, we employ a Bayesian computational model
\footnote{For a guide on Bayesian techniques, see \cite{schad2021toward,van2021bayesian,dix2022bayesian, burkner2017brms,kay2016researcher}, we used the tutorial of Nathaniel Haine's as a starting point for our modelling efforts: \hyperlink{}{http://haines-lab.com/post/2021-01-10-modeling-classic-effects-dunning-kruger/}}.
 a hierarchical Bayesian model to jointly estimate participants' objective and perceived performance while accounting for latent skill, metacognitive bias, and metacognitive noise
\footnote{For a theoretical model, see \citet{burson2006skilled}}.
To allow for a baseline comparison of the DKE, we modelled the data jointly with that of \citet{jansen2021rational}, whose study did not involve AI. This approach enables us to compare the DKE in our sample, where participants used AI, with the non-AI sample of \citet{jansen2021rational}. The model accounts for ceiling effects in performance estimates, treating scores of 20 as censored observations.
Specifically, our hierarchical Bayesian model accounts for the presence of AI ($k = \text{"AI"}$ or $\text{"no AI"}$) in estimating both the participants' achieved performance and their estimated performance. It also integrates latent skill, metacognitive bias, and group-level metacognitive noise, which scales the bias and latent skill. Metacognitive bias, in the model, reflects each person's over- or underestimation of their skill, while metacognitive noise reflects the lack of information regarding their own skill.



Let \( \theta_i \) represent the relative latent skill for participant \( i \) with the prior \( \theta_i  \sim \mathcal{N}(0, 2) \).

The objective performance \( y_{\text{obj},i} \) is modeled as:
\[
y_{\text{obj},i} \sim \text{Binomial}\left(n_{\text{obj}}, \Phi_{\text{approx}}(\theta_i)\right),
\]
where \( n_{\text{obj}} \) is the total number of items and \( \Phi_{\text{approx}}(\cdot) \) is the approximation for the cumulative standard normal distribution. 

Perceived performance \( y_{\text{per},i} \) is influenced by group level bias \( b_k \), latent skill \( \theta_i \) and noise \( \sigma_k \), which scales the difference of bias \( b_k \) and latent skill \( \theta_i \):

\[
y_{\text{per},i} \sim 
\begin{cases} 
\text{Binomial}\left(n_{\text{per}}, \Phi_{\text{approx}}\left(\frac{ \theta_i + b_k}{\sigma_k}\right)\right), & y_{\text{per},i} < n_{\text{per}}, \\
\text{Censored-Binomial}\left(n_{\text{per}}, \Phi_{\text{approx}}\left(\frac{\theta_i + b_k}{\sigma_k}\right)\right), & y_{\text{per},i} = n_{\text{per}}.
\end{cases}
\]

Here, \( n_{\text{per}} \) is the total number of perceived items. The priors for bias \( b_k \) and noise \(\sigma_k\) are the following:

\[
b_k \sim \mathcal{N}(0, 2), \quad \sigma_k \sim \text{LogNormal}(0, 2).
\]

\begin{figure}
    \centering
    \includegraphics[width=1\linewidth]{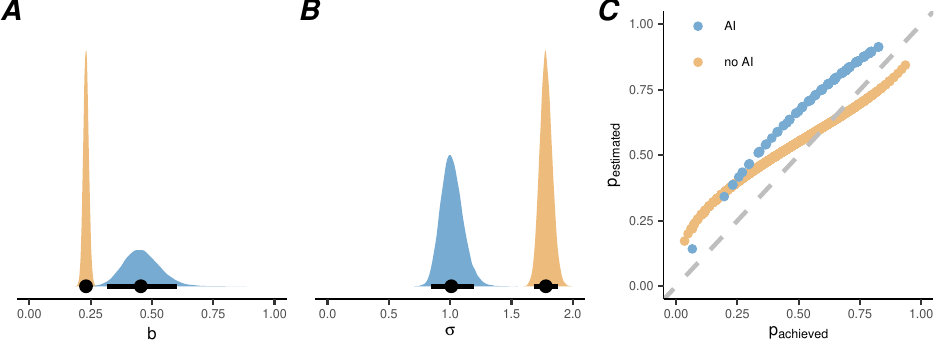}
    \caption{Comparison posterior distributions with median and 95\% HDI for the model parameters $b$ for bias in each group for bias (Plot A)  and 
    $\sigma$  (Plot B). The posterior distributions of the AI group (in blue) and no AI group (in yellow).  Plot C shows the average posterior predicted values for percent correct achieved (x-axis) and percent correct expected (y-axis) for each group. The s-shape around ideal metacognitive accuracy (grey line) indicates a DKE with low-performers overestimating their performance more than high-performers (yellow; no AI group). }
    \Description{Plot (a): comparison of posterior distributions for the bias parameter between the AI group and no-AI group (larger for the AI group). Plot (b): comparison of posterior distributions of metacognitive noise for AI and non-AI groups, where the AI group shows less noise in judgement than the no-AI group. Plot (c): scatter plot showing the posterior predicted values for percent correct achieved (x-axis) and percent correct expected (y-axis) for each group. The s-shape around ideal metacognitive accuracy indicates a DKE with low-performers overestimating their performance more than high-performers.}
    \label{fig:compmodelpar}
  
\end{figure}

Our model mitigates the issue around regression to the mean by explicitly modelling latent skill \( \theta_i \) as a continuous variable with a flexible distribution. By incorporating noise \( \sigma_k \) that scales skill level \( \theta_i \) and bias \( b_k \), the model allows for greater variability in judgment among low-skill participants. This scaling effect, coupled with a bias term \( b_k \) and hierarchical priors, reduces the tendency for all participants to regress toward a single mean. 
For a DKE to exist, bias that is \( b_k \) > 0 and noise that is \(  \sigma_k \) > 1 has to be satisfied. If only metacognitive bias is driving a DKE pattern, then \( \sigma_k \) will be centered at 1 (values of \( \sigma_k \) under 1 and close to zero would mean that high-performers would be less accurate in estimating their performance
\footnote{To fit our data into the model, we used the STAN-sampler \cite{carpenter2017stan}. Four Hamilton-Monte-Carlo chains were computed, each with 15,000 iterations and a 30\% warm-up. Trace plots of the Markov-chain Monte-Carlo permutations were inspected for divergent transitions and autocorrelation, and we checked for local convergence. All Rubin-Gelman statistics \cite{gelman1992inference} were well below 1.1, and the Effective Sampling Size was over 1000. 

We then analyzed the posterior of the model. To investigate a parameter's distinguishability from zero, we utilized $p_b$, which resembles the classical $p$-value but quantifies the effect's likelihood of being zero (for $b$) and one (for $\sigma$) or opposite \cite{hoijtink2018testing,shi2020reconnecting}. Effects with $p_b$ $\leq$ 2.5\% were deemed distinguishable. We also calculated the 95\% High-Density Interval (HDI) for each model parameter; for visualization of prior and posterior, see \autoref{fig:compmodelpar}.}).
The bias parameter for our AI-interaction sample, \(b_{AI}\), showed a median of 0.45 (95\% HDI [0.32, 0.60]). The consistently positive bias indicates that individuals, when using AI, tend to overestimate their abilities. We also find a bias \(b_{no AI}\) for the non-AI group of \citet{jansen2021rational} (Median = 0.23 (95\% HDI [0.21, 0.25], $p_b$ = 0.0\%) although when comparing posterior samples (see \autoref{fig:compmodelpar}), 99\% of posterior samples were larger in the AI group as compared to the non-augmented sample. 

To understand how metacognitive noise affected self-assessment, we can turn to \(\sigma_k\). For the non-AI group, we find a \( \sigma_{no AI} \) above 1, indicating noise affecting self-assessment. This group had a median of 1.78 (95\% HDI [1.69, 1.88], $p_b$ = 0.0\%, see \autoref{fig:compmodelpar}B), indicating noise in judgment for the sample of \citet{jansen2021rational}. Combined with bias, this contributes to the DKE (see also \autoref{fig:compmodelpar}C for posterior predictions from the model). In comparison, our sample, which used AI to complete the task, was not affected by the DKE (Median =  1.01, 95\% HDI [0.84, 1.19], $p_b$ = 45.66\%). Given the non-overlapping distributions (0\% overlap) and the small HDI's, we can assert that with \( \sigma_{AI} \) being over 1, scaling of the equation of bias and skill is not present in our sample. Hence, when augmented with AI, we observe no DKE (see again \autoref{fig:compmodelpar}C). 


\subsubsection{Qualitative Data}

In addition to quantitative measurements, we analyzed the qualitative data collected during Study 1 using an inductive thematic approach \cite{clarke2017thematic}. This included both the prompts participants entered into the AI chatbot interface and their responses to an open-ended question at the end of the questionnaire. 
The prompts were filtered to exclude those that were a direct copy from the task, ensuring that only meaningful interactions were kept for this analysis. The remaining prompts were then inductively analyzed to identify recurring themes (section \ref{prompt_analysis}). Responses to the open-ended question were analyzed to explore differences in AI perception during the interactions, where recurrent themes were found (section \ref{open_end_qs}).

\subsubsection{Analysis of prompts}\label{prompt_analysis}
In our study, we collected 11944 prompts from 246 participants, each answering 20 logical reasoning questions using the AI chatbot. To ensure quality of the data, we employed a filtering process to exclude prompts that appeared to be direct copies of questions and options verbatim, identifying these through cosine similarity comparisons with the original questions. Prompts with a cosine similarity level exceeding 95\% were likely to be full copies of both the questions and options. These prompts were considered less informative and, therefore, were excluded. Prompts with cosine similarity levels between 60\% and 95\% likely included only the questions without the options, being also removed.

After filtering, the dataset was reduced from 11944 to 3074 distinct prompts, encompassing a variety of interactions with the AI such as salutations, queries about specific response options, requests for further clarifications, assessment of AI's capabilities, inquiries regarding the model's confidence in its responses, and other task-related instructions and inquiries.


\paragraph{Analysis of open-ended questions}\label{open_end_qs}

A qualitative analysis of the open-ended question "Please describe how you used the AI Chatbot" revealed diverse types of perceptions and interactions with AI among participants, highlighting varying degrees of user reliance, collaboration, and trust. 

\begin{table}[ht]
\caption{Participant Approaches to AI Use in Study 1}
\centering
\label{tab:qualtable}
\begin{tabular}{p{2.5cm} p{6cm}ll}
\hline
\textbf{Category} & \textbf{Description} & \textbf{Actual (\textit{M ± SD}}) & \textbf{Perceived (\textit{M ± SD})} \\ \hline
High level of trust & Participants relied heavily on AI ("blindly trusted"), copying and pasting questions without critically assessing AI's outputs or further inquiry. 58.94\% (145 out of 246)  & 13.014 ± 3.062 & 16.844 ± 3.679 \\ \hline
Collaborative Partner & Participants perceived AI as a collaborative partner rather than a mere tool, engaging in joint problem-solving and using inclusive language ("we did this") when describing their interactions.  12.60\% (31 out of 246) & 13.065 ± 3.176 & 18.161 ± 1.881 \\ \hline
Complementary Tool & Participants used AI strictly as a complementary tool for verification of their independently formulated answers, maintaining control of the process. 21.54\% (53 out of 246) & 10.5 ± 3.317 & 16.000 ± 6.733 \\ \hline
Inconclusive/Did Not Use & Participants either provided inconclusive responses regarding the strategy used during the experience or did not find the AI tool useful. 6.91\% (17 out of 246) & 13.302 ± 2.729 & 16.302 ± 4.012 \\ 
\end{tabular}
\Description{Different categories of participant approaches to AI use based on trust, collaboration, and reliance. Four categories are listed: High level of trust, Collaborative Partner, Complementary Tool and Inconclusive/Did Not Use. Each category provides mean and standard deviation (M ± SD) for both actual and perceived performance. }
\end{table}

The majority of participants demonstrated a high level of trust in AI, often accepting its suggestions without further inquiry. This behavior raises concerns about overreliance on AI, as noted by \citet{lu2021human}. 
12.60\% of participants perceived AI as a collaborative partner, using inclusive language and viewing it as part of a joint effort rather than just a tool.
Another 21.54\% of participants viewed AI strictly as a complementary tool, using it cautiously for verification while maintaining control of the problem-solving process.  
Finally, 6.5\% either provided inconclusive responses regarding the strategy used or did not find the AI tool useful. 
The data reveals diverse ways participants perceived AI, providing insights into HAI dynamics and individual variability.

\subsection{Interim Dicussion}
We found that using ChatGPT augmented our sample beyond a no-AI benchmark (i.e. AI augmentation) but that only a little more than half of our sample could surpass the AI alone (i.e. human-AI synergy). We found that most people overestimated their performance with AI and that there was no indication of a DKE when using the AI system. This may be due to participants' tendency to not reflect on their performance and low metacognitive sensitivity, which is corroborated by our qualitative reports of people copying and pasting questions to the chat interfaces and then taking the AI's answer without reflection.

\section{Study 2: Incentivizing metacognitive thinking}

To address the potential confound of a lack of motivation in our sample to engage in metacognitive monitoring, we conducted a second study in which participants received a monetary incentive for accurate judgments across the task (for a DKE study employing incentives, see \citet{ehrlinger2008unskilled} Study 3). If participants monitor their performance when incentivized, the DKE could resurface. 
Given that \citet{jansen2021rational} did not incentivize participants for accurate metacognitive judgments, this also mandated the sampling of a no-AI group within our study setup. We thus sampled another 250 participants for each the AI and the no-AI group to see if an incentive can motivate metacognitive monitoring and analyze the quantitative data. All data and analysis scripts for Study 2  can be found at \osfrepository.



\subsection{Method}





We recruited 500 English-speaking participants located in the USA through Prolific. The sample was split into two groups: 250 participants completed the task without AI assistance (no-AI group) and 250 participants completed the task with AI assistance (AI group). 
Participants solved the same 20 logical reasoning problems used in Study 1. We did not collect the SNAIL for the no-AI group.  Study 2 took, on average, 25 minutes to complete for the no-AI group and around 52 minutes for the AI group. Each participant was compensated 7 pounds per hour.

To motivate accurate self-assessment, participants in both groups were informed they would receive monetary compensation based on the accuracy of their performance estimates (compensation would be given to participants whose presumed number of correct answers closely matched their actual score). This incentive aimed to motivate participants to engage critically with the task and closely monitor their performance. All participants received full benefit compensation of 0.50 pounds regardless of their achieved performance.
Similarly to Study 1, we included an attention check where participants were required to read a brief study and task description. They then answered two multiple-choice questions, one about the topic (logical reasoning) and another regarding how they could receive additional compensation (good judgment).

From 500 participants, we analyzed 452 participants (age: $M=37.24, SD=13.36$): 245 in the AI group and 207 in the no-AI group. Across both groups, 48 participants were excluded (3 for missing data, 3 for invalid performance estimates (exceeding 20 correct answers), and 42 for too low completion times. 
 
202 participants identified as female, 242 as male, 6 as non-binary, 1 as two-spirit, and 1 who preferred not to disclose. Their highest educational degrees included 21 doctoral, 132 Master’s-level, 194 Bachelor’s-level, 64 upper secondary school, and 41 vocational qualifications. Regarding English proficiency, 391 participants identified as native speakers, 56 as fully fluent, 4 as conversationally fluent, and 1 as having basic proficiency.

For the AI group, a subset of participants (26) reported prior experience taking the LSAT. Their performance in this study ($M= 13.50, SD = 1.98$) was comparable to those without prior LSAT experience (n = 219; $M = 13.25, SD = 2.55$), and they were not excluded from the analysis.
For the no-AI group, 25 participants reported prior LSAT experience, performing similarly ($M = 9.50, SD = 4.06$) to those without LSAT experience (n = 182; $M = 9.52, SD = 3.60$).

\subsection{Results and Discussion}

\begin{figure}
    \centering
    \includegraphics[width=\linewidth]{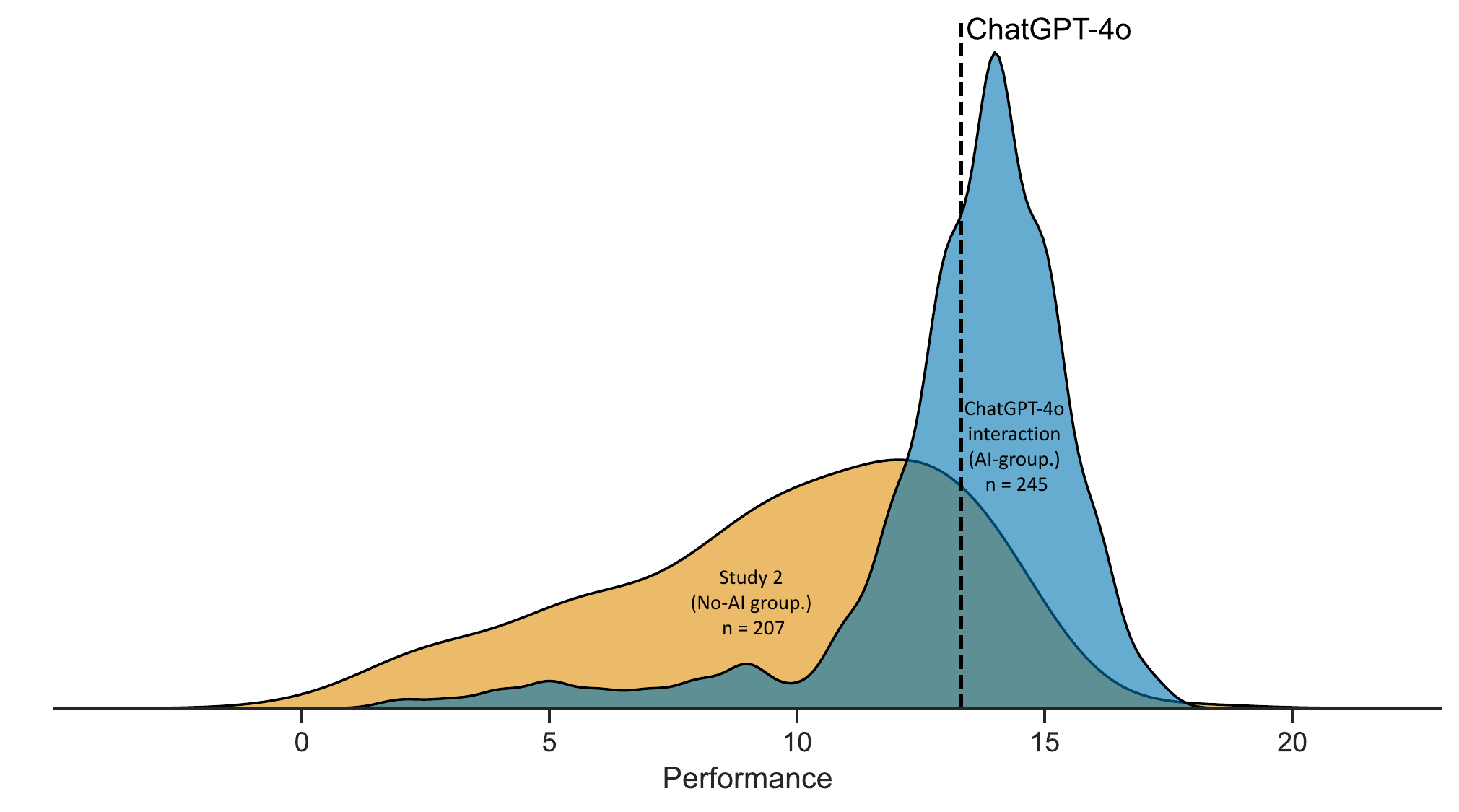}
    \caption{Comparison of performance scores between the sample of participants interacting with ChatGPT and the sample of participants without AI. The blue density curve represents the distribution of performance scores in a sample of 245 participants using ChatGPT, showing a peak in performance at around 14 points. The yellow density curve corresponds to the sample of participants in the No-AI group, with lower overall performance scores. The vertical dashed line indicates the mean performance score in the ChatGPT simulation.}
    \label{fig:synergyperformance_newstudy}
    \Description{The graph compares performance scores between participants using ChatGPT and those without AI assistance from study 2. The ChatGPT group shows higher performance, peaking around 14 points, while the no-AI group has lower overall scores. The dashed line marks the mean performance in the ChatGPT simulation.}
\end{figure}

Participants in the AI group performed on average slightly worse as compared to AI alone ($M$ = 13.31, $SD$ =  2.44, $t(244) = -2.17, p = .031, d = -0.14$) but better than the no-AI group ($M$ = 9.71, $SD$ =  3.59, $t(450) = 12.60, p < .001, d = 1.18$). Therefore, we can assert that, on average, using AI has augmented performance but not that there is a synergy effect. In the AI group, 59.18\% of participants scored higher than ChatGPT, with a total of 145 participants out of 245 surpassing its performance. In the no-AI group, 14.49\% of participants scored higher than ChatGPT, corresponding to 30 participants out of 207, see also \autoref{fig:synergyperformance_newstudy}. Therefore, performance in Study 2 mirrors Study 1.

\begin{figure}[!h]
    \centering
    \begin{subfigure}[b]{0.49\linewidth}
        \centering
        \includegraphics[width=\linewidth]{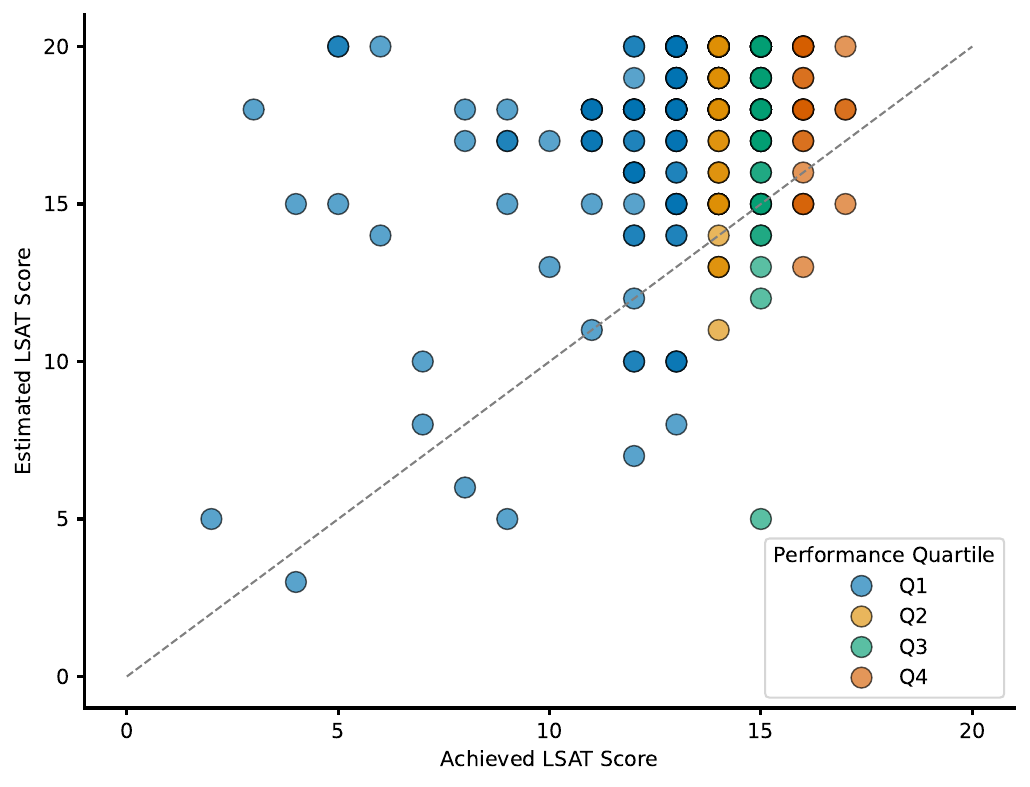}
        \caption{Scatter-plot of Estimated LSAT Scores as a function of Achieved LSAT Scores for the AI group in Study 2.}
        \label{fig:corplotS2}
        \Description{}
    \end{subfigure}
    \hfill
    \begin{subfigure}[b]{0.49\linewidth}
        \centering
        \includegraphics[width=\linewidth]{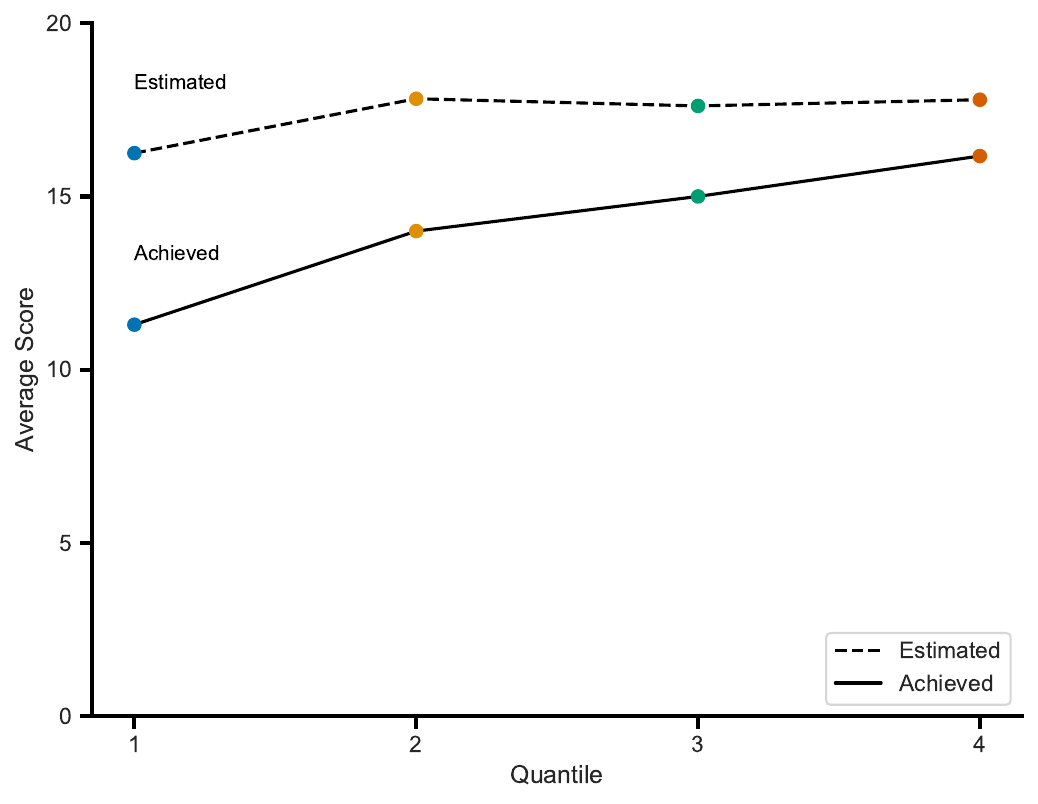}
        \caption{Classical Dunning-Kruger Quartile-plot. Comparison of Mean Estimated and Mean Achieved Scores Across Quartiles for the AI group in Study 2.}
        \label{fig:dkplotS2}
        \Description{}
    \end{subfigure}
    \caption{Correlation of estimated and achieved LSAT score from different perspectives, individual \autoref{fig:corplotS2} vs. quartile-level \autoref{fig:dkplotS2} for the AI group in Study 2.}
    \Description{Plot (a): scatter plot of estimated LSAT scores as a function of achieved LSAT scores across quartiles (Q1 to Q4). While some participants were accurate in their estimates, some participants were very wrong in estimating their performance accurately. Plot (b): quartile-level Dunning-Kruger plot, showing that estimated scores are consistently higher than achieved scores across all quartiles.}
    \label{fig:combinedS2}
    
\end{figure}

Investigating metacognitive accuracy for each sample, we find that in the AI group, participants overestimate their performance ($M$ = 17.13, $SD$ =3.16) which differed significantly from zero when subtracting individual performance ($t(244) = 18.33, p < .001, d = 1.17$). The same was found for the no-AI group ($M$ = 13.62, $SD$ = 4.14, $t(206) = 11.81, p < .001, d = 0.82$). Both quantile's metacognitive accuracy differed from 0 for AI (Q1: $t(99) = -12.78, p < .001, d = -1.28$, Q4: $t(23) = -3.83, p < .001, d = -0.78$) as well as the no-AI group (Q1: $t(51) = -10.73, p < .001, d = -1.49$, Q4: $t(51) = -3.43, p = .001, d = -0.48$). Comparing estimates of estimated performance and performance of the first and the fourth quartile for each group, we find that the lowest quartile overestimates their performance relatively more when compared to the best-performing quartile (AI: $t(122) = 4.06, p < .001, d = 0.73$, no-AI: $t(102) = 7.25, p < .001, d = 1.42$), see \autoref{fig:combinedS2}. Therefore, the pattern of results in Study 2 regarding metacognitive accuracy also closely resembles Study 1. 
Comparing the confidence for correct and incorrect responses for the AI group, we find that, on average, participants are more confident for correct ($M$ = 85.95, $SD$ = 13.71) as compared to incorrect responses ($M$ = 82.57, $SD$ = 15.61;$t(244) = 5.39, p < .001, d = 0.34$). While confidence was descriptively lower for the no-AI group, we find the same pattern  (correct: $M$ = 77.57, $SD$ = 14.85; incorrect: $M$ = 73.04, $SD$ = 16.73; $t(205) = 6.33, p < .001, d = 0.44$). A slight increase in confidence when accurate metacognition is incentivized is consistent with \citet{ehrlinger2008unskilled}.

Conducting a ROC analysis for each group, we found that most participants could distinguish between correct and incorrect answers. The mean AUC for the AI group ($M$ = .62, $SD$ =0.12) and the no-AI group ($M$ = .61, $SD$ = 0.11) differed from .5 (AI: $t(244) = 15.30, p < .001, d = 0.98$; no-AI: $t(205) = 14.26, p < .001, d = 0.99$ ), with most people exceeding the threshold of .5 AUC (AI: 196 of 245; no-AI: 172 of 207).
The pattern of correlations of performance, metacognitive measures and AI literacy also resembled Study 1 in the AI group, see \autoref{tab:CorrelationsS2}, and in the no-AI group, see \autoref{tab:CorrelationssS2noAI}. 

\begin{table}
\caption{Correlation Table of Metacognitive measures and AI literacy as measured by the SNAIL in Study 2}
\label{tab:CorrelationsS2}

\centering
\begin{tabular}[t]{llllllllll}
\toprule
  & $\Delta EP$ & Estimate & Performance & $\Delta conf$ & $\mu conf$ & AUC & SNAIL TU & SNAIL CA & SNAIL PA\\
\midrule
$\Delta EP$ &  &  &  &  &  &  &  &  & \\
Estimate & 0.71*** &  &  &  &  &  &  &  & \\
Performance & -0.42*** & 0.34*** &  &  &  &  &  &  & \\
$\Delta conf$ & -0.05 & -0.05 & -0.01 &  &  &  &  &  & \\
$\mu conf$ & 0.39*** & 0.62*** & 0.28*** & -0.09 &  &  &  &  & \\
AUC & -0.16* & -0.09 & 0.09 & 0.49*** & -0.23*** &  &  &  & \\
SNAIL TU & 0.20** & 0.18** & -0.04 & -0.26*** & 0.21** & -0.21*** &  &  & \\
SNAIL CA & 0.14* & 0.21*** & 0.08 & -0.08 & 0.14* & -0.12 & 0.68*** &  & \\
SNAIL PA & 0.09 & 0.19** & 0.12 & -0.13* & 0.22*** & -0.15* & 0.75*** & 0.83*** & \\
\bottomrule
\end{tabular}
\noindent \raggedright \textbf{Note}. $df$ = 243, * indicates $p$ < .05,** indicates $p$ < .01 and *** indicates $p$ < .001. $\Delta EP$ represents the difference between performance and estimated performance (metacognitive accuracy). Performance refers to the achieved task performance. $\Delta conf$ is the difference between predicted and actual confidence, while $\mu conf$ is the mean confidence (average confidence ratings). AUC refers to Area Under the Curve, with a higher AUC value indicating a more reliable confidence score in reflecting participants' correctness; SNAIL TU stands for the Technical Understanding score, SNAIL CA represents the Critical Appraisal score, and SNAIL PA is the Practical Application score.
\Description{Correlations between metacognitive measures and AI literacy (SNAIL). Key variables include the difference between estimated and actual performance, confidence measures, Area Under the Curve, and three SNAIL components: Technical Understanding score (SNAIL TU), Critical Appraisal score (SNAIL CA), and Practical Application score (SNAIL PA)}
\end{table}

\begin{figure}
    \centering
    \includegraphics[width=1\linewidth]{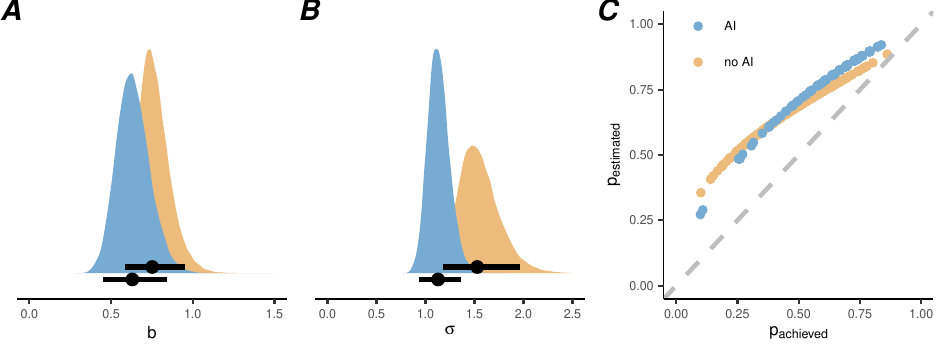}
    \caption{Comparison posterior distributions with median and 95\% HDI for the model parameters $b$ for bias in each group for bias (Plot A)  and 
    $\sigma$  (Plot B) for the second study. The posterior distributions of the AI group (in blue) and no AI group (in yellow).  Plot C shows the average posterior predicted values for percent correct achieved (x-axis) and percent correct expected (y-axis) for each group. The s-shape around ideal metacognitive accuracy (grey line) indicates a DKE with low-performers overestimating their performance more than high-performers (yellow; no AI group). }
    \Description{Plot (a): comparison of posterior distributions for the bias parameter between the AI group and no-AI group (larger for AI group). Plot (b): comparison of posterior distributions of metacognitive noise for AI and non-AI groups, where the AI group shows less noise in judgement than the no-AI group. Plot (c): scatter plot showing the posterior predicted values for percent correct achieved (x-axis) and percent correct expected (y-axis) for each group. The s-shape around ideal metacognitive accuracy indicates a DKE with low-performers overestimating their performance more than high-performers.}
    \label{fig:compmodelpar_newstudy}
  
\end{figure}

Applying our computational model using the same priors and sampler configuration as in Study 1, we find that both AI (Median = 0.63 (95\% HDI [0.45, 0.85], $p_b$ = 0.0\%) and no-AI (Median = 0.75 (95\% HDI [0.58, 0.95], $p_b$ = 0.0\%) show a metacognitive bias without distinguishing clearly between the AI group as compared to the non-AI sample;  18.1\% of posterior samples were larger in the AI group as compared to the no-AI group, see also \autoref{fig:compmodelpar_newstudy}A. Note that the discrepancy to Study 1, likely comes from the relatively lower precision, given the smaller sample size in Study 2. \(\sigma_k\) indicating metacognitive noise was found to be above 1 for the no-AI group (Median = 1.53 (95\% HDI [1.18, 1.96], $p_b$ = 0.1\%), resembling the DKE pattern in Study 1, but centred around 1 for the AI group (Median = 1.13 (95\% HDI [0.93, 1.36], $p_b$ = 10.1\%), see also see also \autoref{fig:compmodelpar_newstudy}B. 2.46\% of posterior samples for sigma in the no-AI group exceed the AI group. Therefore, metacognitive noise does not scale the bias for the AI group, but it does for the no-AI group. We replicate the pattern of results in Study 1 again; see also \autoref{fig:compmodelpar_newstudy}C. The difference in shape for \autoref{fig:compmodelpar_newstudy}C and \autoref{fig:compmodelpar}C can be explained by the difference in range, especially regarding high performance, see \autoref{fig:synergyperformance} and compare to \autoref{fig:synergyperformance_newstudy}.

Overall, we can replicate the results of Study 1 in Study 2. Giving an incentive for accurate metacognitive judgements did not activate a DKE pattern for participants using AI. Nevertheless, we can see that the absolute levels of performance overestimation are slightly larger for the no-AI group in our sample (e.g., comparing estimated performance across studies).


\section{General Discussion}

This paper offers insights into metacognitive monitoring in HAI by examining how users with varying competence interacted with AI during logical reasoning tasks. We explored the impact of AI on metacognitive accuracy, focusing on the DKE, user confidence, and AI literacy in two studies. Our findings reveal a significant inability to assess one’s performance accurately when using AI equally across our sample.

\subsection{Effect of AI Literacy on Metacognition in Human-AI Interaction}

While AI users in our sample outperformed those in \citet{jansen2021rational}, they consistently overestimated their performance by about four points, aligning with previous HCI research \cite{kosch2023placebo,villa2023placebo,kloft2023ai}. The moderate correlation between estimated and actual scores (\autoref{tab:Correlations}), with many participants estimating their joint performance with AI higher than the most skilled in the sample (\autoref{fig:corplot}), suggests that AI improves performance but leads to highly biased self-assessments. 

This disconnect between actual and perceived performance mirrors earlier findings on overtrust and overreliance in AI systems \cite{kloft2023ai,shekar2024people,lu2021human}. Overconfidence may impair users’ ability to evaluate their performance without AI, posing challenges for designing balanced human-AI interfaces. 
The classic DKE, where lower performers overestimate and higher performers underestimate their performance, disappeared with AI use, suggesting AI levels performance but does not correct inflated self-assessments. Metacognitive bias was doubled for the entire sample compared to \citet{jansen2021rational} in Study 1, which was less pronounced in Study 2.

We have also found an unexpected link between AI literacy and metacognitive accuracy across both studies. Participants with higher AI literacy were less accurate in self-assessments, contradicting the assumption that AI literacy improves metacognitive monitoring. Familiarity with AI may enhance the better-than-average effect \cite{brown1986evaluations,zell2020better}, leading to overestimation of both relative and absolute performance. 

Metacognitive sensitivity further explains these effects. Our ROC analysis showed that while participants were generally confident, they tended to overestimate the correctness of incorrect responses, indicating low metacognitive sensitivity. This suggests participants did not adequately monitor their performance when using AI and were largely unaware of their performance post-decision.

Qualitative data revealed varied perceptions of AI’s role, from a tool to a teammate, but these differences did not descriptively show an effect on performance or metacognitive accuracy, see \autoref{tab:qualtable}, contradicting theories that suggest interaction framing impacts outcomes \cite{pataranutaporn_influencing_2023, villa2023placebo}. Regardless of user perception, the core metacognitive challenges in HAI persist.

\subsection{Integration into theory}

For HCI, high metacognitive bias leads users to overestimate their performance and over-rely on AI systems \cite{ma_are_2024}, reducing their ability to critically monitor HAI outcomes \cite{tankelevitch2023metacognitive}.
From a computational rationality perspective \cite{oulasvirta2022computational}, this bias may be an adaptive response to AI presence, as participants may optimize perceived utility (e.g., efficiency) rather than monitoring HAI. This is supported by low metacognitive sensitivity and participants’ reliance on copy-pasting rather than higher-level metacognitive strategies (e.g., AI as collaborator). Such aligns with \citet{villa2023placebo}, who found reduced error-processing when participants believed they were using sham AI. Therefore, our study provides further evidence of diminished metacognitive monitoring in HAI.

A lack of HAI monitoring explains several effects. First, it clarifies why AI use leads to adverse learning outcomes \cite{abbas2024harmful,bastani2024generative}; users are overly optimistic and fail to monitor evolving joint performance. Second, while AI tools offer perceived empowerment and efficiency \cite{kloft2023ai,kosch2023placebo,villa2023placebo}, the lack of reflection hinders users' ability to assess real benefits (placebo effect). Additionally, it explains persistent overreliance and overtrust \cite{klingbeil2024trust}, and why AI explanations are rarely integrated into behavior \cite{bansal_does_2021,wang2021explanations,ghassemi2021false}.
Though psychology and HCI offer methods to improve metacognitive judgments (refer to \autoref{tab:designprinciples}), they may provide only temporary solutions. \citet{rafner2022deskilling} calls for a systemic, long-term strategy, considering cognitive and metacognitive deskilling risks at individual, team, and organizational levels. This emphasizes the need for strategies that foster cognitive resilience and critical engagement with AI over short-term fixes targeting immediate metacognitive deficits.

\begin{table}[htbp]
\centering
\caption{Issues, Consequences, and Design Principles to Address Impaired Metacognitive Monitoring in Human-AI Interaction}
\label{tab:designprinciples}
\begin{tabular}{p{3cm} p{4cm} p{7cm}}
\hline
\textbf{Metacognitive Issue} & \textbf{Consequences} & \textbf{Design Principles} \\ 
\hline
\textbf{Overreliance on AI outputs} 
& Users trust AI outputs without critical assessment, leading to reduced self-reflection and failure to notice AI errors.
& \begin{itemize}
    \item Confidence calibration to align user confidence with AI output uncertainty \cite{ma_are_2024}
    \item AI uncertainty visualization to make AI output reliability transparent \cite{beauxis2021role,prabhudesai2023understanding}
    \item Explanatory AI interfaces to clarify AI decision-making processes and enable users to assess validity \cite{karran2022designing}
\end{itemize} \\ 
\hline
\textbf{Loss of metacognitive monitoring} 
& Users are unable to accurately assess their own performance or monitor task progress, especially in complex decision-making tasks.
& \begin{itemize}
    \item Post-task reflection to encourage users to evaluate their performance after interacting with AI (for a starting point, see \cite{tankelevitch2023metacognitive})
    \item Cognitive forcing strategies such as prompts to promote critical thinking and reduce automatic reliance on AI outputs \cite{forcing2021}
\end{itemize} \\
\hline
\end{tabular}
\Description{Metacognitive issues, consequences, and design principles to address impaired monitoring in human-AI interaction. Two issues are listed: Overreliance on AI outputs and loss of metacognitive monitoring. Design principles include confidence calibration, AI uncertainty visualization, explanatory AI interfaces, post-task reflection, and cognitive forcing strategies to encourage critical thinking and reduce overreliance.}
\end{table}

\subsection{Limitations}

While our study provides valuable insights into metacognitive monitoring in HAI, several limitations may affect the generalizability and interpretation of our findings.
First, the apparent absence of the DKE ("being unskilled and unaware" \cite{dunning2011dunning}), in our study may not fully reflect the underlying cognitive dynamics. AI interaction may have made participants equally skilled yet unaware of their performance rather than eliminating the DKE. AI-enhanced performance might have decoupled metacognitive judgments from cognitive performance (e.g., high confidence, low sensitivity, and high bias).

Second, we relied on LSAT questions as our logical reasoning task. While these assess logical reasoning, they may not capture the diversity of real-world reasoning skills or domains. Furthermore, these tasks might overlap with the AI’s training data, limiting generalizability. However, since ChatGPT-4o's performance was imperfect (68.25\%), we believe our findings still apply to other tasks.

Third, our focus on LSAT-based reasoning limits the scope of metacognitive biases across domains. Future research should use diverse tasks (e.g., writing creative texts with an LLM) to examine how AI interaction affects metacognitive monitoring and whether improvements in accuracy generalize across tasks.
Fourth, we found little effect of different AI strategies on metacognitive accuracy and performance. The AI’s role, whether as a tool, collaborator, or teammate, did not impact participants' performance evaluation (see again \autoref{tab:qualtable}). Future research should explore how different AI roles (e.g., tool vs. collaborator) influence metacognitive accuracy and task performance.

Fifth, AI literacy was self-reported, which raises the same challenges inherent to the DKE: individuals with limited knowledge may overestimate their expertise. In line with this, self-reported AI literacy might not even directly translate into effective interactions with AI. Thus, how AI literacy links to performance and metacognition deserves to be further explored in future studies. 

Sixth, participants were required to prompt ChatGPT at least once and then proceed with their decision, which may not reflect naturalistic usage patterns. In real-world scenarios, users might consult AI tools multiple times, ignore them entirely, or encounter more proactive AI systems offering suggestions unsolicited. Future studies should examine how varying degrees of user engagement with the AI and the AI’s proactiveness influence metacognitive accuracy, sensitivity, and bias. 

Finally, exploring long-term learning scenarios—such as those examined in \citet{bastani2024generative}—could illuminate how metacognitive processes evolve as individuals repeatedly interact with AI systems. Over extended periods, accurate metacognitive monitoring may become increasingly vital for achieving sustained performance gains (e.g., learning calculus with AI assistance and then taking an exam); our study cannot speak to the role of metacognition in AI-mediated learning contexts.

\subsection{Implications}

Following \citet{van2023implications}, we identify three types of implications of our work: theoretical, design and methodological.
For HCI theory, while biases in human-AI interaction have been studied \citep{liu2019well,kloft2023ai,bertrand2022cognitive,haliburton2024investigating}, little research has analyzed the metacognitive mechanisms underlying these biases and their impact on interaction \cite{tankelevitch2023metacognitive}. Applying metacognition theory to HAI (e.g., see  \citet{colombatto2023illusions}), offers a new perspective for improving metacognitive monitoring in interaction design. Our findings also suggest that AI literacy alone is insufficient for optimal metacognitive abilities in HAI. 
For design, we propose a new HAI interaction model that integrates metacognition and new design concepts. In the short term, designers should adjust interaction models, similar to \citet{forcing2021}, and consider broader sociotechnical risks \cite{rafner2022deskilling}. 
Furthermore, real-time sensing of metacognitive states (e.g., user agency) and adapting interfaces to cognitive states \cite{chiossi2023adapting,evalai2024} could be crucial for optimizing HAI.
Methodologically, our study introduces a new approach to analyze biased decision-making in repeated AI interactions. Using computational methods, researchers can distinguish general biases from those emerging in post-decision processes (e.g., increased metacognitive noise), allowing for a more precise analysis of bias development in HAI.

\section{Conclusion}


We found that participants using an LLM to complete a logical reasoning task were unable to accurately assess their own performance, often overestimating it. With AI use, the DKE was eliminated. Based on these findings, we suggest developing new interfaces for interactive AI that are designed to enhance metacognition, allowing users to monitor their performance more accurately.


\begin{acks}
To Robert, for the bagels and explaining CMYK and color spaces.
\end{acks}

\bibliographystyle{ACM-Reference-Format}
\bibliography{hybIntel}

\appendix
\newpage
\section{Tables}

\begin{table}[hbp]
\caption{Correlation Table of Metacognitive Measures in Study 2 for the no-AI group}
\label{tab:CorrelationssS2noAI}

\centering
\begin{tabular}[t]{lllllll}
\toprule
  & $\Delta EP$ & Estimate & Performance & $\Delta conf$ & $\mu conf$ & AUC \\
\midrule
$\Delta EP$ &  &  &  &  &  & \\
Estimate & 0.68*** &  &  &  &  & \\
Performance & -0.54*** & 0.25*** &  &  &  & \\
$\Delta conf$ & -0.22** & -0.15* & 0.11 &  &  & \\
$\mu conf$ & 0.38*** & 0.60*** & 0.20** & -0.18** &  & \\
AUC & -0.12 & -0.13 & 0.01 & 0.57*** & -0.24*** & \\
\bottomrule
\end{tabular}

\noindent \raggedright \textbf{Note.} $df = 205$, * $p < .05$, ** $p < .01$, *** $p < .001$. $\Delta EP$ represents the difference between performance and estimated performance (metacognitive accuracy). Performance refers to the achieved task performance. $\Delta conf$ is the difference between predicted and actual confidence, while $\mu conf$ is the mean confidence (average confidence ratings). AUC refers to Area Under the Curve, with a higher AUC value indicating a more reliable confidence score in reflecting participants' correctness.
\end{table}

\section{Figures}
\begin{figure}[!h]
    \centering
    \begin{subfigure}[b]{0.49\linewidth}
        \centering
        \includegraphics[width=\linewidth]{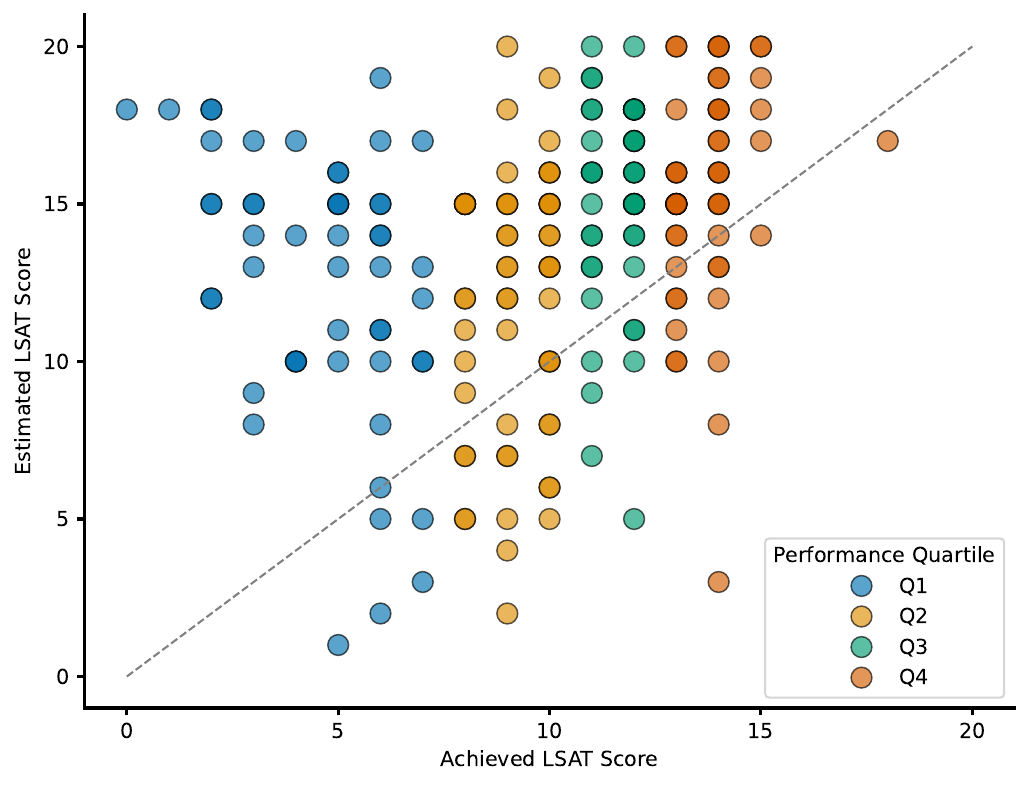}
        \caption{Scatter-plot of Estimated LSAT Scores as a function of Achieved LSAT Scores for the no-AI group in Study 2.}
        \label{fig:corplot_baseline}
        \Description{}
    \end{subfigure}
    \hfill
    \begin{subfigure}[b]{0.48\linewidth}
        \centering
        \includegraphics[width=\linewidth]{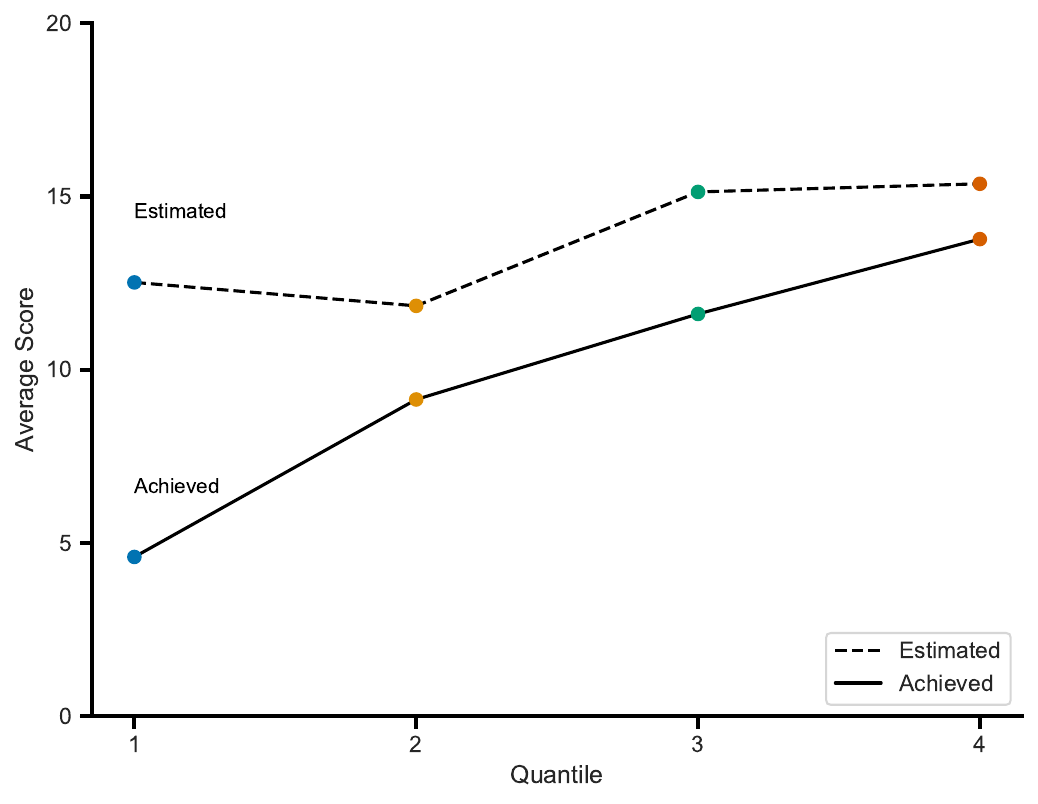}
        \caption{Classical Dunning-Kruger Quartile-plot. Comparison of Mean Estimated and Mean Achieved Scores Across Quartiles for the no-AI group in Study 2.}
        \label{fig:dkplot_baseline}
        \Description{}
    \end{subfigure}
    \caption{Correlation of estimated and achieved LSAT score from different perspectives, individual \autoref{fig:corplot_baseline} vs. quartile-level \autoref{fig:dkplot_baseline} for the no-AI group in Study 2.}
    \Description{Plot (a): scatter plot of estimated LSAT scores as a function of achieved LSAT scores across quartiles (Q1 to Q4) in the no-AI group in study 2. While some participants were accurate in their estimates, some participants were very wrong in estimating their performance accurately. Plot (b): quartile-level Dunning-Kruger plot for the no-AI group of study 2, showing that estimated scores are consistently higher than achieved scores across all quartiles.}
    \label{fig:combined_baseline}
    
\end{figure} 



\end{document}